\documentclass[review]{elsarticle}

\usepackage{lineno,hyperref}
\modulolinenumbers[5]
\usepackage{amssymb}
\usepackage{underscore}
\usepackage{indentfirst}
\usepackage{amsmath,amsthm}
\usepackage{txfonts}
\usepackage{subfigure}
\usepackage{epsfig}
\usepackage{float}
\usepackage{algorithm}
\usepackage{algpseudocode}
\usepackage{booktabs,graphicx}
\usepackage{caption}
\usepackage{enumitem}
\usepackage{graphicx}
\usepackage{times}  
\usepackage{helvet}  
\usepackage{courier}  
\usepackage{url}  
\usepackage{graphicx}  
\usepackage{subfigure}
\usepackage[section]{placeins}
\usepackage{float}
\usepackage{multirow}
\usepackage{textcomp}
\usepackage{tabularx}
\usepackage{amsthm,amsmath,amsfonts,amssymb}
\usepackage{amsmath,epsfig,amssymb,subfigure,bm,dsfont}
\usepackage{siunitx}
\usepackage{array}
\usepackage{amsmath,amssymb,amsfonts}
\usepackage{graphicx}
\usepackage{subfigure}
\usepackage[section]{placeins}
\usepackage{float}
\usepackage{multirow}
\usepackage{textcomp}
\usepackage{algpseudocode}
\usepackage{amsmath}
\usepackage{graphics}
\usepackage{epsfig}
\usepackage{bm}
\usepackage{threeparttable}
\usepackage{epstopdf}
\usepackage{color}

\newcommand{\PreserveBackslash}[1]{\let\temp=\\#1\let\\=\temp}
\newcolumntype{C}[1]{>{\PreserveBackslash\centering}p{#1}}
\newcolumntype{R}[1]{>{\PreserveBackslash\raggedleft}p{#1}}
\newcolumntype{L}[1]{>{\PreserveBackslash\raggedright}p{#1}}

\journal{Knowledge-Based Systems}

\begin{document}

\begin{frontmatter}

\title{Task-adaptive Asymmetric Deep Cross-modal Hashing}

\author[mymainaddress]{Fengling Li}
\author[mymainaddress]{Tong Wang}
\author[mymainaddress]{Lei Zhu}
\address[mymainaddress]{School of Information Science and Engineering, Shandong Normal University }
\author[mymainaddress1]{ Zheng Zhang}
\address[mymainaddress1]{Bio-Computing Research Center,Harbin Institute of Technology }
\author[mymainaddress]{Xinhua Wang}
\ead{wangxinhua@sdnu.edu.cn}

%
%

\begin{abstract}
Supervised cross-modal hashing aims to embed the semantic correlations of heterogeneous modality data into the binary hash codes with discriminative semantic labels. It can support efficient large-scale cross-modal retrieval due to the fast retrieval speed and low storage cost. However, existing methods equally handle the cross-modal retrieval tasks, and simply learn the same couple of hash functions in a symmetric way. Under such circumstances, the characteristics of different cross-modal retrieval tasks are ignored and sub-optimal performance may be brought. Motivated by this, we present a Task-adaptive Asymmetric Deep Cross-modal Hashing (TA-ADCMH) method in this paper. It can learn task-adaptive hash functions for two sub-retrieval tasks via simultaneous modality representation and asymmetric hash learning. Different from previous cross-modal hashing methods, our learning framework jointly optimizes the semantic preserving from multi-modal features to the hash codes, and the semantic regression from query modality representation to the explicit labels. With our model, the learned hash codes can effectively preserve the multi-modal semantic correlations, and meanwhile, adaptively capture the query semantics. Besides, we design an efficient discrete optimization strategy to directly learn the binary hash codes, which alleviates the relaxing quantization errors.
Extensive experiments demonstrate the state-of-the-art performance of the proposed TA-ADCMH from various aspects.
\end{abstract}

\end{frontmatter}


\section{Introduction}
\label{intro}
Cross-modal retrieval \cite{aaa1} takes a certain kind of modality data as query objects to retrieve the relevant data in other modalities. Meanwhile, with the development of multimedia technology, large amounts of heterogeneous multi-modal data are explosively generated. To tackle the retrieval efficiency problem, cross-modal hashing \cite{ACQ,zheng2016hetero,cao2018cross,liu2018fast} is proposed to project the high-dimensional multi-modal data into the low-dimensional binary hash codes, which are forced to express consistent semantics with the original data.
For the high retrieval and storage efficiency, cross-modal hashing has aroused considerable attention to solve large-scale cross-modal retrieval.

With the trend, many cross-modal hashing methods are proposed in literature.
Generally, they can be coarsely divided into two categories:
unsupervised \cite{ij:conf/ijcai/KumarU11,IEEEexample:inter-media,IEEEexample:linear,IEEEexample:cmfh,LIONG2018114,8594588} and supervised \cite{zhang2014large,IEEEexample:ann2,tang2016supervised,wang2015semantic,IEEEexample:ann6,wang2018label} cross-modal hashing. Unsupervised cross-modal hashing methods learn the low-dimensional embedding of original data without any semantic labels.
The learned hash codes try to maintain the semantic relevance of heterogeneous multi-modal data. Contrastively, supervised cross-modal hashing methods learn more discriminative hash codes with the supervision of explicit semantic labels.

Most existing cross-modal hashing methods are based on the shallow models. They usually adopt linear or simple non-linear mapping as the hash function, which may limit the discriminative capability of modality feature representation, and thus degrade the
retrieval accuracy of the learned hash codes. Recently, deep cross-modal hashing \cite{zhang2018unsupervised,hu2018deep,jiang2017deep,zhong2018deep} is proposed to simultaneously perform deep representation and hash code learning. They replace the linear mapping with multi-layer nonlinear mapping and thus capture the intrinsic semantic correlations of cross-modal instances more effectively. It has been proved that cross-modal hashing methods based on deep models have better performance than the shallow hash methods which directly adopt hand-crafted features.

 Although great successes have been achieved by existing cross-modal hashing methods, they equally handle the cross-modal retrieval tasks (e.g. image retrieves text and text retrieves image), and simply learn the same couple of hash functions for them.
 Under such circumstance,
 the characteristics of different cross-modal retrieval tasks are ignored, which may lead to suboptimal performance. To tackle the limitations, this paper proposes a Task-adaptive Asymmetric Deep Cross-modal Hashing (TA-ADCMH) method to learn task-specific hash functions for each cross-modal sub-retrieval tasks. The main contributions and innovations of this paper are stated as follows:
\begin{itemize}
  \item  We propose a new supervised asymmetric hash learning framework based on deep neural networks for large-scale cross-modal retrieval. Two couples of deep hash functions can be learned for different cross-modal retrieval tasks by performing simultaneous deep feature representation and asymmetric hash learning.
      To the best of our knowledge, there is still no similar work.

   \item In asymmetric hash learning part,
   we preserve the semantic similarity across different modalities and enhance the representation capability of query modality.
   With such design, the learned hash codes can capture the semantic correlations of multi-modal data, as well as the query semantics of the specific cross-modal retrieval task.
  \item An iterative optimization algorithm is proposed to enable the discreteness of hash codes and alleviate the errors of binary quantization.
  Experimental results demonstrate the superiority of the proposed method on three public cross-modal retrieval datasets.
\end{itemize}


\section{Literature review of cross-modal hashing and multi-task learning}
\subsection{Unsupervised cross-modal hashing}
 Unsupervised cross-modal hashing transforms the modality features into the shared hash codes by preserving the original similarities. Representative works include Cross-View Hashing (CVH) \cite{ij:conf/ijcai/KumarU11}, Inter-Media Hashing (IMH) \cite{IEEEexample:inter-media}, Linear Cross-Modal Hashing (LCMH) \cite{IEEEexample:linear}, Collective Matrix Factorization Hashing (CMFH) \cite{IEEEexample:cmfh}, Latent Semantic Sparse Hashing (LSSH) \cite{Zhou:2014:LSS:2600428.2609610}, Robust and Flexible Discrete Hashing (RFDH) \cite{wang2017robust}, Cross-Modal Discrete Hashing (CMDH) \cite{LIONG2018114} and Collective Reconstructive Embeddings (CRE) \cite{8594588}. CVH is a typical graph-based hashing method, which is extended from the standard spectral hashing \cite{weiss2009spectral}.
 It transforms the original multi-view data into the binary codes by minimizing the weighted Hamming distances. IMH maps heterogeneous multimedia data into hash codes by constructing graphs. It learns the hash functions by linear regression for new instances. Its joint learning scheme can effectively preserve the inter- and intra- modality consistency. LCMH first leverages k-means clustering to represent each training sample as a k-dimensional vector, and then maps the vector into the to-be-learnt binary codes. CMFH utilizes the collective matrix factorization model to transform multimedia data into
 the low-dimensional latent representations and then approximates them with hash codes.
 It also fuses the multi-view information to enhance the retrieval accuracy. LSSH follows similar idea of CMFH. It attempts to learn the latent factor matrix for image structure by sparse coding and text concepts by matrix decomposing. Compared with CMFH, it can better capture high-level semantic correlation for similarity search across different modalities. RFDH first learns unified hash codes for each training data by employing discrete collaborative matrix factorization. Then, it jointly adopts $l_{2,1}$-norm and adaptive weight of each modality to enhance the robustness and flexibility of hash codes. CMDH proposes a discrete optimization strategy to learn the unified binary codes for multiple modalities. This strategy projects the heterogeneous data into a low-dimensional latent semantic space by using matrix factorization. The latent features are quantified as the hash codes by projection matrix. CRE is proposed to learn unified binary codes and binary mappings for different modalities by collective reconstructive embedding. It bridges the semantic gap between heterogeneous data.

\subsection{Supervised cross-modal hashing}
Supervised cross-modal hashing generates the hash codes under the guidance of semantic information. Typical methods include Semantic Correlation Maximization (SCM) \cite{zhang2014large}, Semantics-Preserving Hashing (SePH) \cite{IEEEexample:ann2}, Supervised Matrix Factorization Hashing (SMFH) \cite{tang2016supervised}, Semantic Topic Multimodal Hashing (STMH) \cite{wang2015semantic}, Discrete Latent Factor Model based Cross-Modal Hashing (DLFH) \cite{jiang2019discrete}, Discrete Cross-modal Hashing (DCH) \cite{IEEEexample:ann6} and Label Consistent Matrix Factorization Hashing (LCMFH) \cite{wang2018label}. SCM aims at preserving maximum semantic information into hash codes by avoiding computing pair-wise semantic matrix explicitly. It improves both the retrieval speed and space utilization. SePH first employs probability distribution to preserve supervision information of multi-modal data, and then the hash codes can be obtained by solving the problem of Kullback-Leibler divergence. SMFH is developed based on the collective matrix factorization. It employs graph Laplacian \cite{TNNLSciteleizhu2020,TNNLS2019citeleizhu,TCYB2019citeleizhu,TCYB2018citeleizhu} and semantic labels to learn binary codes for multi-modal data. STMH employs semantic modeling to detect different semantic themes for texts and images respectively, and then maps the captured semantic representations into a low-dimensional latent space to obtain hash codes. DLFH proposes an efficient hash learning algorithm based on the discrete latent factor model to directly learn binary hash codes for cross-modal retrieval. DCH is an extended application of Supervised Discrete Hashing (SDH) \cite{shen2015supervised} in multi-modal retrieval. It learns a set of modality-dependence hash projections as well as discriminative binary codes to keep the classification consistent with the label for multi-modal data. LCMFH leverages the auxiliary matrix to project the original multi-modal data to the low-dimensional representation of latent space, and quantizes it with semantic labels to the hash codes.

All of the above hashing methods are shallow methods, which construct hash functions by simple linear or nonlinear transformation.
Thus, these methods cannot effectively explore the semantic correlations of heterogeneous multi-modal data.

\subsection{Deep cross-modal hashing}
Deep cross-modal hashing basically seeks a common binary semantic space via multi-layer nonlinear projection from multiple heterogeneous modalities. State-of-the-art deep cross-modal hashing methods include Unsupervised Generative Adversarial Cross-modal Hashing (UGACH) \cite{zhang2018unsupervised},
Deep Binary Reconstruction for Cross-modal Hashing (DBRC) \cite{hu2018deep},
Deep Cross-Modal Hashing (DCMH) \cite{jiang2017deep},
Discrete Deep Cross-Modal Hashing (DDCMH) \cite{zhong2018deep},
Self-Supervised Adversarial Hashing (SSAH) \cite{li2018self}
and Deep Joint-Semantics Reconstructing Hashing (DJSRH) \cite{DJSRH2019}.
UGACH uses the generation model and discriminant model to promote the learning of hash function.
It incorporates the correlation graph into the learning procedure to capture the intrinsic manifold structures of multi-modal data. DBRC develops a deep network based on a special Multimodal Restricted Boltzmann Machine (MRBM) to learn binary codes. The network employs the adaptive tanh hash function to obtain the binary valued representation instead of joint real value representation, and reconstructs the original data to preserve the maximum semantic similarity across different modalities.
DCMH first extracts the deep features of text and image modalities through two neural networks,
and then preserves the similarity of two different deep features
into the unified hash codes by using a pair-wise similarity matrix.
DDCMH proposes a cross-modal deep neural network to directly encode the binary hash codes by employing discrete optimization, which can effectively preserve the intra- and inter-modality semantic correlation. SSAH devices a deep self-supervised adversarial network to solve cross-modal hashing problem. This network combines multi-label semantic information and adversarial learning to eliminate the semantic gap between deep features extracted from heterogeneous modalities.
DJSRH constructs a novel joint-semantics affinity matrix which integrates the original neighborhood from different modalities and trains the networks to generate binary codes that maximally reconstruct the joint-semantics relations via the proposed reconstructing framework.

\emph{Differences}: The existing deep cross-modal hashing approaches equally handle the different cross-modal retrieval tasks when constructing the hash functions.
Under such circumstances, the characteristics of cross-modal retrieval tasks are ignored during the hash learning process and it might lead to sub optimal performance.
Different from them, in our paper, we propose a task-adaptive cross-modal hash learning model to learn two couples of hash functions for two cross-modal sub-retrieval tasks respectively. In our model, the semantic similarity across different modalities are preserved and the representation capability of query modality is enhanced. With the learning framework, the learned hash codes can simultaneously capture the semantic correlation of different modalities and the query semantics of the specific cross-modal retrieval task.
\subsection{Multi-task learning}
Multi-Task Learning (MTL) \cite{MTL} is a technology that includes multiple related tasks, each of which can be a general learning task.
It is found that learning these tasks jointly can bring performance improvement instead of learning them individually.
Inspired by the success of multi-task learning in various aspects, many new techniques have been proposed.
Co-Attentive Multi-task Learning (CAML) \cite{MTL1} proposes a co-attentive multi-task learning model for explainable recommendation by fully mining the correlations between the recommendation task and the explanation task.
Deng et al. \cite{MTL2} proposes a novel multi-task learning scheme that leverages multi-view attention mechanism to interact with different tasks and learn more comprehensive sentence representations.
Multi-Task Field-weighted Factorization Machine (MT-FwFM) \cite{MTL3} formulates conversion prediction as a multi-task learning problem
and jointly learns the prediction models for different types of conversions.
The feature representations shared by all tasks and model-specific parameters provide the benefits of information-sharing across all tasks.
Multi-task learning-based Unsupervised Domain Adaptation (mtUDA) \cite{MTL4} relaxes the single classifier assumption
in the conventional classifier-based unsupervised domain adaptation and proposes to jointly optimize source and target classifiers
by considering the manifold structure of the target domain and the distribution divergence between the domains.
Multi-Context Embedding Network (MCENet) \cite{MTL5} proposes a multi-context representation approach for object counting.
It extracts the potential features for the appearance context and the semantic context by the first subnetwork and
transfers the learned features into the two parallel and complementary subnetworks.
Thus the multiple contexts are represented and embedded to assist the counting task.
Different from these multi-task learning schemes, in this paper,
we propose a new supervised asymmetric hash learning framework based on deep neural networks for large-scale cross-modal retrieval.
It can learn task-adaptive hash functions for sub-retrieval tasks (image retrieves text or text retrieves image) via simultaneous modality representation and asymmetric hash learning.
To the best of our knowledge, there is still no similar work.
\label{sec:3}
\label{sec:3.1}
\captionsetup[table]{labelfont={bf},name={Table },singlelinecheck=off,font={small}}
\begin{table}[t]
\caption{The list of main notations. }
\label{tab:1}
\centering
\setlength{\tabcolsep}{0.04\linewidth}{
\begin{tabular}{ll}
\hline\noalign{\smallskip}
Notation & Description\\
\noalign{\smallskip}\hline\noalign{\smallskip}
$\textbf{X} $ &the raw image matrix \\
$\textbf{Y}\in \mathbb{R}^{n \times {d_y}}$ & the text feature matrix \\
$\textbf{F}\in \mathbb{R}^{r\times n}$&deep feature representation matrix of image  \\
$\textbf{G}\in \mathbb{R}^{r\times n}$&deep feature representation matrix of text \\
$\textbf{P}\in \mathbb{R}^{r\times c}$&semantic projection matrix of image\\
$\textbf{W}\in \mathbb{R}^{r\times c}$&semantic projection matrix of text\\
$\textbf{S}\in \mathbb{R}^{n\times n}$ & pair-wise semantic matrix\\
$\textbf{L}\in \mathbb{R}^{c\times n}$&  point-wise semantic label\\
$\textbf{B}_t\in \mathbb{R}^{r\times n}$&binary hash codes  \\
$N_t$&mini-batch size \\
\textit{$d_y$}&the dimension of text \\
\textit{c}&the number of classes\\
\textit{r}& hash code length\\%
\textit{t}&the number of retrieval tasks \\
\noalign{\smallskip}\hline
\end{tabular}}
\end{table}
\section{Task-adaptive asymmetric deep cross-modal hashing}
\subsection{Notations and problem definition}
Assuming that a database with $n$ training instances denoted as $\textbf{O}=\left\{{\textbf{o}}_{i}\right\}_{i=1}^n$, each training instance $\textbf{o}_{i}$ is comprised of two modalities: image and text. $\textbf{X}={[\textbf{x}_1,\textbf{x}_2,...,\textbf{x}_n]^\texttt{T}}$ denotes the raw image matrix. $\textbf{Y}=[\textbf{y}_1,\textbf{y}_2,...,\textbf{y}_n]^\texttt{T}\in\mathbb{R}^{n \times {d_y}}$ represents the text feature matrix with $d_y$ dimensions. Each instance $\textbf{y}_{i}$ is associated with an instance $\textbf{x}_{i}$. Besides, the point-wise semantic label is given as $\textbf{L} = [\textbf{l}_1,\textbf{l}_2,...,\textbf{l}_n]\in\mathbb{R}^{c\times {n}}$, where $c$ is the total number of categories and $l_{ki}= 1$ implies that $\textbf{o}_i$ belongs to class $k$, otherwise $l_{ki}= 0$. We define the pairwise semantic matrix $\textbf{S}\in\mathbb{R}^{n \times n}$
with its $(i,j)$-th element denoted as $S_{ij}\in\left\{0,1\right\}$.
$S_{ij}=1$ indicates that the image $\textbf{x}_i$ is similar to the text $\textbf{y}_i$ while $S_{ij}=0$ means they are dissimilar to each other. In general, cross-modal retrieval problem (includes two modalities image and text) has two sub-retrieval tasks:
one task is the image retrieving text (I2T),
and the other task is text retrieving image (T2I). The goal of our method is to learn two nonlinear hash functions $h^{(x)}(x)$ and $h^{(y)}(y)$ for different cross-modal retrieval tasks, the binary hash codes $\textbf{B}_1$ relates to images hash functions $h^{(x)}(x)$ for I2T task, and the binary hash codes $\textbf{B}_2$ relates to texts hash functions $h^{(y)}(y)$ for T2I task. $r$ is the length of hash codes, and $t$ is the number of retrieval tasks.
The main notations used in the paper are listed in Table \ref{tab:1}.

\subsection{Framework overview}
\begin{figure*}[t]
\centering
\subfigure{\includegraphics[scale=0.38]{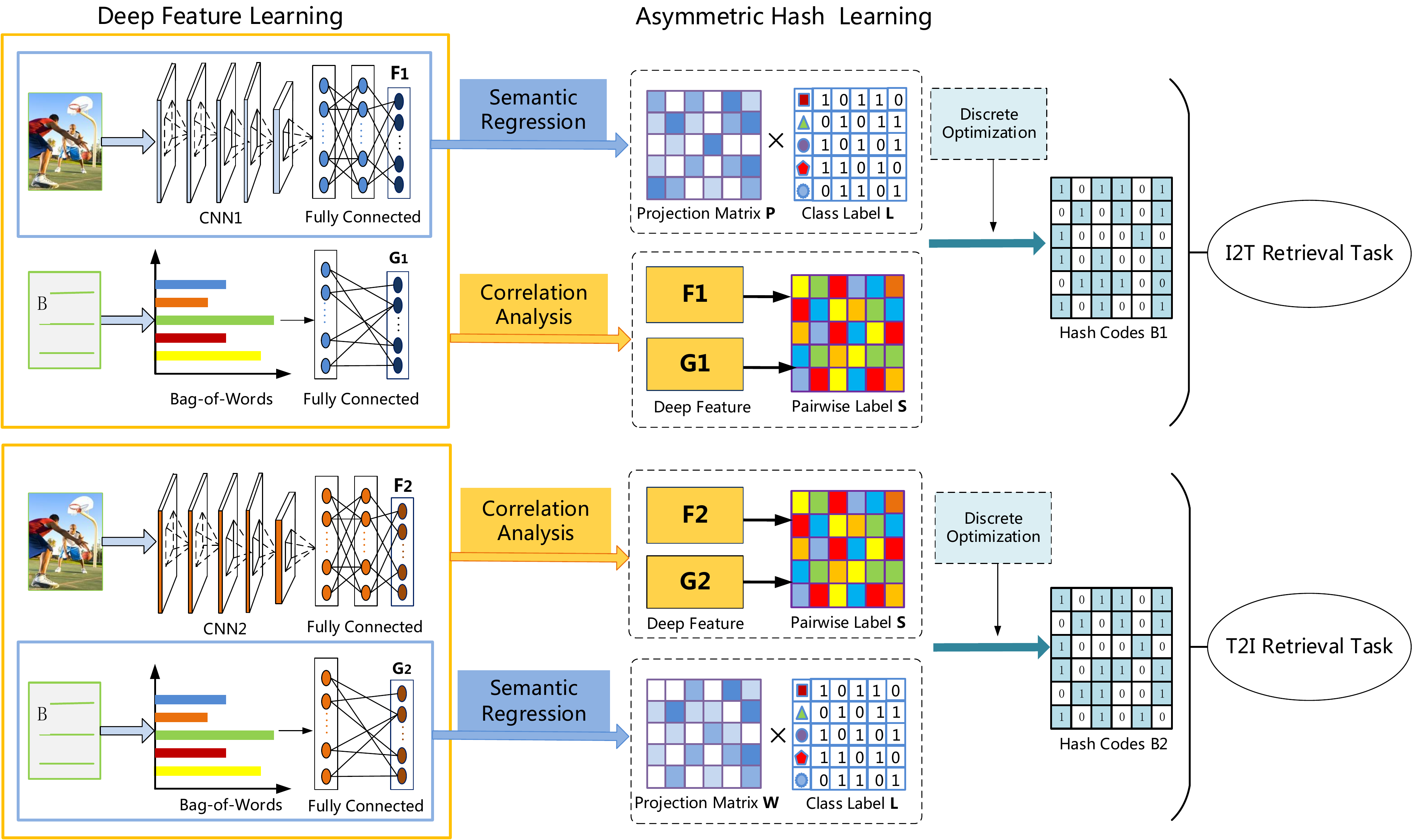}}
\caption{The overall learning framework of our TA-ADCMH method. }
\label{Figfram}
\end{figure*}

The basic framework of the proposed TA-ADCMH method is illustrated in Figure \ref{Figfram}.
 The framework illustrates the cross-modal retrieval tasks into two sub-retrieval tasks: I2T and T2I. Each retrieval task consists of two main components: deep feature learning and asymmetric hash learning. In deep feature learning part, an image-text pair is imported into the deep neural networks of  I2T and T2I retrieval tasks respectively, and two couples of deep models are trained to extract corresponding feature representations for the images and texts. In asymmetric hash learning part, we learn two sets hash codes by jointly performing the semantic regression and correlation analysis for two retrieval tasks. Semantic regression of two retrieval tasks uses the point-wise label, and correlation analysis uses the pair-wise label. Semantic regression is represented by blue color, while the correlation analysis is represented by orange color.
Specifically, (a) we perform the semantic regression from deep image feature to point-wise label to classify the images of different classes with the learned representation. (b) The correlation analysis leverages pairwise label to narrow semantic gap between image and text modalities. (c) We perform the semantic regression from deep text feature to point-wise label to classify texts of different classes with the learned representation. With the efficient discrete optimization, we learn the discriminative hash codes for different retrieval tasks.

\subsection{Model formulation}
In this paper, we propose a supervised asymmetric deep cross-modal hashing model to independently perform two different cross-modal retrieval tasks, each of which mainly includes two parts: deep feature learning and asymmetric hash learning.
In the first part,
we extract the deep feature representations of image and text by two couples of deep neural networks. In the second part, we perform asymmetric hash learning to capture the semantic correlations of multimedia data with the supervision of pair-wise semantic matrix and enhance the discriminative capability of query modality representation with point-wise semantic label.
\subsubsection{Deep feature learning}
 In the deep feature learning part, we design two couples of deep neural networks for two cross-modal sub-retrieval tasks. As shown in Figure \ref{Figfram}, we can find that each pair of image-text deep networks are used to perform I2T and T2I sub-retrieval tasks, respectively. To be fair, we use similar deep neural networks of image modality for two sub-retrieval tasks. Both two deep networks are based on convolutional neural network (CNN-F) \cite{CNNF}
 and initialized with the neural network weights which are pre-trained on the large-scale ImageNet dataset \cite{deng2009imagenet}. Particularly, CNN-F is an eight layer deep network structure with five convolution layers (conv1 - conv5) and three fully-connected (Fc6_I - Fc8_I) layers. We modify the structure of last fully-connected layer
 by setting the number of hidden units as the length of hash codes, and adopt identity function as the activation function for the last network layer. We also use two deep neural networks of text modality for two sub-retrieval tasks, each of which consists of two fully-connected layers (Fc6_T - Fc7_T). Particularly, we represent the original text vectors as the Bag-of-Words (BOW) \cite{yang2015robust} which is then used as the input to the deep neural network. Further, we obtain the hash codes as the outputs from the last fully-connected layer. Similar to the image network, we also adopt identity function as the activation function. In this paper, $\emph{f}_t(\textbf{X}_i;{\theta_{x_t}})|_{t=1}^2$ represents the image feature representation model, which is determined as the outputs
of the last fully-connected layer with data $\textbf{x}_i$, where $\theta_{x_t}|_{t=1}^2$ is the weight parameters of deep image neural networks. $\emph{g}_t(\textbf{Y}_j;{\theta_{y_t}})|_{t=1}^2$ represents the text feature representation model, which is determined as the outputs
of the last fully-connected layer with data $\textbf{y}_j$, where $\theta_{y_t}|_{t=1}^2$ is the weight parameters of deep text neural networks. We iteratively optimize the deep neural networks of image and text modalities, and simultaneously obtain deep hash functions and deep features from the last layer of fully-connected layer. The deep feature representations will be further approximated to hash codes. This part integrates both feature representation learning and hash codes learning, which enables the energy of the learned compact feature to be well preserved in the binary codes.

\subsubsection{Asymmetric hash learning for I2T}
 The cross-modal retrieval task concentrates on two sub-retrieval tasks: image retrieves text and text retrieves image. Previous methods generally learn the same couple of hash functions in a symmetric way for two different cross-modal retrieval tasks.
 They cannot effectively capture the query semantics during the non-linear multi-modal mapping process,
 as the characteristics of different cross-modal retrieval tasks are ignored.
 To address these problems, in this paper, we develop an asymmetric hash learning model to learn different hash functions for different retrieval tasks. Specifically, for each task, besides optimizing the semantic preserving of multi-modal data into hash codes, we perform the semantic regression from query-specific modality representation to the explicit labels. With such design, the semantic correlations of multi-modal data can be preserved into the hash codes, and simultaneously, the query semantics can be captured adaptively.

The overall objective function of I2T sub-retrieval task is formulated as
\begin{equation}
\label{eq:W}
\begin{split}
\min \limits_{\mathbf{B}_1,\theta_{x_1},\theta_{y_1},\textbf{P}} \mathcal{J}_1 =-&\underbrace{\sum\limits_{i,j=1}^n(S_{ij}\Phi_{ij}\,-\, \log(1+e^{\Phi_{ij}}))+\lambda_1\lVert{\textbf{B}_1-\textbf{F}_1\rVert}_F^2+\beta_1\lVert{\textbf{B}_1-\textbf{G}_1\rVert}_F^2}_{\mbox{ Correlation\;  Analysis} } \\
&+\underbrace{\mu_1\lVert{\textbf{F}_1-\textbf{P}\textbf{L}\rVert}_F^2+\nu_1\mathnormal{R}(\textbf{F}_1\textbf{1},\textbf{G}_1\textbf{1},\textbf{P})}_{\mbox{ Semantic\;  Regression} } \\
 &\quad s.t. \, \textbf{B}_1\in{\left \{-1,1 \right \}}^{r\times n}\\
\end{split}
\end{equation}
where $\lambda_1$, $\beta_1$, $\mu_1$, $\nu_1$ are the regularization parameters, $\Phi_{ij}= \frac{1}{2}\textbf{F}_{1i}^\texttt{T}\textbf{G}_{1j}$, $\textbf{F}_1 \in\mathbb{R}^{r\times n}$ with $\emph{f}_1(\textbf{X}_i;{\theta_{x_1}})$, $\textbf{G}_1\in\mathbb{R}^{r\times n}$ with $\emph{g}_1(\textbf{Y}_j;{\theta_{y_1}})$ are the deep features extracted from images and texts respectively. $\mathbf{B}_{1}\in\mathbb{R}^{r\times n}$ is the binary hash codes to be learned for I2T task
and we impose discrete constraint to obtain binary values.
$\textbf{L} = [\textbf{l}_1,\textbf{l}_2,...,\textbf{l}_n]$ is the point-wise semantic label. $\textbf{P}\in\mathbb{R}^{r\times c}$ is the semantic projection matrix which supports the semantic regression from image (query) modality representation $\textbf{F}_1$ to \textbf{L}.
The first term in Eq.(\ref{eq:W}) is a negative log likelihood function, which is based on the likelihood function defined as
\begin{eqnarray}
p(S_{ij}|\textbf{F}_{1i},\,\textbf{G}_{1j})\ =\
\begin{cases}
\;\sigma(\Phi_{ij})\ &\ \text{$S_{ij}$\ =\ 1}\\
\;1-\sigma(\Phi_{ij})\ &\ \text{$S_{ij}$\ =\ 0}
\end{cases}
\end{eqnarray}
where $\sigma(\Phi_{ij})=\frac{1}{1+e^{-\Phi_{ij}}}$. The negative log likelihood function can make $\textbf{F}_{1i}$ and $\textbf{G}_{1j}$ as similar as possible when $S_{ij}=1$, and be dissimilar when $S_{ij}=0$. Thus, this term can preserve the semantic correlation between deep image feature $\textbf{F}_{1i}$ and deep text feature $\textbf{G}_{1j}$ by the pair-wise semantic supervision. The second and third terms in Eq.(\ref{eq:W}) transform the deep features $\textbf{F}_{1i}$ and $\textbf{G}_{1j}$ into the binary hash codes $\textbf{B}_1$, which collectively preserve the cross-modal semantics into the binary hash codes.
The last term is to avoid overfitting. It is defined as below:
  \begin{equation}
\label{eq:33}
\begin{split}
\nu_1\mathnormal{R}( \, \textbf{F}_1\textbf{1}, \,\textbf{G}_1\textbf{1}, \,\textbf{P} \,)=\nu_1\,\lVert{\textbf{F}_1\textbf{1}\rVert}_F^2
+\nu_1\,\lVert{\textbf{G}_1\textbf{1}\rVert}_F^2+\nu_1\,\lVert{\textbf{P}\rVert}_F^2
 \end{split}
\end{equation}
 The term $ \nu_1\,\lVert{\textbf{F}_1\textbf{1}\rVert}_F^2
+\nu_1\,\lVert{\textbf{G}_1\textbf{1}\rVert}_F^2\,(\textbf{1}=[1,...,1]^\texttt{T}\in \mathbb{R}^{n})$ is to equally partition the information of each bit and ensure the maximum semantic similarity preserved into hash codes.
\subsubsection{Asymmetric hash learning for T2I}
Different from I2T sub-retrieval task, we directly regress the deep text representation to the corresponding point-wise semantic label to preserve the discriminative information of query modality representation. Specifically, we adopt pair-wise semantic label to learn a new binary hash code $\textbf{B}_2$ to preserve the semantic correlation of multi-modal data and capture the query semantics from texts.

Similar to Eq.(\ref{eq:W}), the objective function for T2I sub-retrieval task is formulated as:

%
\begin{equation}
\label{eq:w2}
\begin{split}
\min \limits_{\mathbf{B}_2,\theta_{x_2},\theta_{y_2},\textbf{W}}\mathcal{J}_2= -&\underbrace{\sum\limits_{i,j=1}^n(S_{ij}\Phi_{ij}\,-\, \log(1+e^{\Phi_{ij}}))+\lambda_2\lVert{\textbf{B}_2-\textbf{F}_2\rVert}_F^2+\beta_2\lVert{\textbf{B}_2-\textbf{G}_2\rVert}_F^2}_{\mbox{ Correlation\;  Analysis} } \\
&+\underbrace{\mu_2\lVert{\textbf{G}_2-\textbf{W}\textbf{L}\rVert}_F^2+\nu_2\mathnormal{R}(\textbf{F}_2\textbf{1},\textbf{G}_2\textbf{1},\textbf{W})}_{\mbox{ Semantic\;  Regression} } \\\
 &\quad s.t. \, B_2\in{\left \{-1,1 \right \}}^{r\times n}\\
\end{split}
\end{equation}
where $\Phi_{ij}=\frac{1}{2}\textbf{F}_{2i}^\texttt{T}\textbf{G}_{2j}$, $\textbf{F}_2 \in\mathbb{R}^{r\times n}$ with $\emph{f}_2(\textbf{X}_2;{\theta_{x_2}})$, $\textbf{G}_2\in\mathbb{R}^{r\times n}$ with $\emph{g}_2(\textbf{Y}_2;{\theta_{y_2}})$ are the deep features extracted from images and
texts respectively. $\textbf{W}\in\mathbb{R}^{r\times c}$ is the semantic projection matrix which supports the semantic regression from text (query) modality representation $\textbf{G}_2$ to \textbf{L}. The balance parameters $\lambda_2$, $\beta_2$, $\mu_2$ and $\nu_2$ are regularization parameters of T2I task. The regularization function $\mathnormal{R}( \, \textbf{F}_2\textbf{2}, \,\textbf{G}_2\textbf{2}, \,\textbf{W} \,)$ is denoted as follows:
\begin{equation}
\label{eq:3}
\begin{split}
\nu_2\mathnormal{R}( \, \textbf{F}_2\textbf{1}, \,\textbf{G}_2\textbf{1}, \,\textbf{W} \,)=\nu_2\,\lVert{\textbf{F}_2\textbf{1}\rVert}_F^2
+\nu_2\,\lVert{\textbf{G}_2\textbf{1}\rVert}_F^2+\nu_2\,\lVert{\textbf{W}\rVert}_F^2
 \end{split}
\end{equation}
This term $ \nu_2\,\lVert{\textbf{F}_2\textbf{1}\rVert}_F^2
+\nu_2\,\lVert{\textbf{G}_2\textbf{1}\rVert}_F^2\,(\textbf{1}=[1,...,1]^\texttt{T}\in \mathbb{R}^{n})$ is same as that for I2T task, which is used to balance each bit of hash codes.
\subsection{Optimization scheme}
The objective functions of I2T and T2I retrieval tasks are all non-convex with the involved variables. In this paper, we propose an iterative optimization method to learn the optimal value for I2T and T2I.

\textbf{Step 1}. Update $\theta_{x}$.

For the I2T sub-retrieval task, the optimization problem for updating $\theta_{x_1}$ in Eq.(\ref{eq:W}):
\begin{equation}
\label{eq:5}
\begin{split}
\min \limits_{\theta_{x_1}}-\sum\limits_{i,j=1}^n(S_{ij}\Phi_{ij}\,-\, &\log(1+e^{\Phi_{ij}}))+\lambda_1\lVert{\textbf{B}_1-\textbf{F}_1\rVert}_F^2
+\mu_1\lVert{\textbf{F}_1-\textbf{P}\textbf{L}\rVert}_F^2+\nu_1\lVert{\textbf{F}_1\textbf{1}\rVert}_F^2\\
\end{split}
\end{equation}
The deep CNN parameter $\theta_{x_1}$ of image modality can be trained by stochastic gradient descent (SGD) \cite{Botte} with the back-propagation (BP) algorithm. In each iteration, we randomly select mini-batch samples from the database to train the network, which relieves the SGD algorithm from falling directly into the local optimal value near the initial point. Specifically, we first compute the following gradient for each instance of $\textbf{x}_i$:
 \begin{equation}
 \label{eq:7}
\begin{split}
 \frac{\partial \mathcal{J}_1}{\partial \textbf{F}_{1i}}=\frac{1}{2}\sum\limits_{j=1}^n(\sigma(\Phi_{ij})\textbf{G}_{1j}-S_{ij}\textbf{G}_{1j})
 +2\lambda_1(\textbf{F}_{1i}-\textbf{B}_{1i})+2\mu_1(\textbf{F}_1-\textbf{P}\textbf{L})+2\nu_1\textbf{F}_1\textbf{1}
\end{split}
\end{equation}
Then we can compute the $\frac{\partial \mathcal{J}_1}{\partial \theta_{x_1}}$ according to the BP updating rule until convergence.

For the T2I retrieval task, the optimization problem for updating $\theta_{x_2}$ in Eq.(\ref{eq:w2}) becomes
\begin{equation}
\label{eq:F}
\begin{split}
 \min \limits_{\theta_{x_2}}-\sum\limits_{i,j=1}^n(S_{ij}\Phi_{ij}\,-\, &\log(1+e^{\Phi_{ij}}))+\lambda_2\lVert{\textbf{B}_2-\textbf{F}_2\rVert}_F^2
+\nu_2\lVert{\textbf{F}_2\textbf{1}\rVert}_F^2\\
\end{split}
\end{equation}
The deep CNN parameter $\theta_{x_2}$ of image modality can be trained by SGD and BP algorithm. Firstly, we compute the following gradient for each instance of $\textbf{x}_i$:
 \begin{equation}
 \label{eq:H}
\begin{split}
 \frac{\partial \mathcal{J}_2}{\partial \textbf{F}_{2i}}=\frac{1}{2}\sum\limits_{j=1}^n(\sigma(\Phi_{ij})\textbf{G}_{2j}-S_{ij}\textbf{G}_{2j})
 +2\lambda_2(\textbf{F}_{2i}-\textbf{B}_{2i})+2\nu_2\textbf{F}_2\textbf{1}
\end{split}
\end{equation}
Then we can compute the $\frac{\partial \mathcal{J_2}}{\partial \theta_{x_2}}$ according to the BP updating rule until convergency.

\textbf{Step 2}. Update $\theta_{y}$.

For the I2T retrieval task, the optimization problem for updating $\theta_{y_1}$ in Eq.(\ref{eq:W}) becomes
\begin{equation}
\label{eq:8}
\begin{split}
 \min \limits_{\theta_{y_1}}-\sum\limits_{i,j=1}^n(S_{ij}\Phi_{ij}\,-\, &\log(1+e^{\Phi_{ij}}))+\beta_1\lVert{\textbf{B}_1-\textbf{G}_1\rVert}_F^2
+\nu_1\lVert{\textbf{G}_1\textbf{1}\rVert}_F^2\\
\end{split}
\end{equation}
The deep CNN parameter $\theta_{y_1}$ of text modality is also trained by SGD and BP algorithm. Firstly, we compute the following gradient for each instance of $y_j$:
 \begin{equation}
 \label{eq:C}
\begin{split}
 \frac{\partial \mathcal{J}_1}{\partial \textbf{G}_{1j}}=\frac{1}{2}\sum\limits_{i=1}^n(\sigma(\Phi_{ij})\textbf{F}_{1i}-S_{ij}\textbf{F}_{1i} ) +2\beta_1(\textbf{G}_{1j}-\textbf{B}_{1j})+2\nu_1\textbf{G}_1\textbf{1}
\end{split}
\end{equation}
Then we can compute the $\frac{\partial \mathcal{J}_1}{\partial \theta_{y_1}}$ according to the BP updating rule until convergency.

For the T2I retrieval task, the optimization for updating $\theta_{y_2}$ in Eq.(\ref{eq:w2}) can be reduced to
\begin{equation}
\label{eq:D}
\begin{split}
\min \limits_{\theta_{y_2}}-\sum\limits_{i,j=1}^n(S_{ij}\Phi_{ij}\,-\, \log(1+e^{\Phi_{ij}}))+\beta_2\lVert{\textbf{B}_2-\textbf{G}_2\rVert}_F^2
+\mu_2\lVert{\textbf{G}_2-\textbf{W}\textbf{L}\rVert}_F^2+\nu_2\lVert{\textbf{G}_2\textbf{1}\rVert}_F^2\\
\end{split}
\end{equation}
The deep CNN parameter $\theta_{y_2}$ of text modality can be learned by SGD and BP algorithm. Firstly, we compute the following gradient for each instance of $\textbf{y}_j$:
 \begin{equation}
 \label{eq:E}
\begin{split}
 \frac{\partial \mathcal{J}_2}{\partial \textbf{G}_{2j}}=\frac{1}{2}\sum\limits_{i=1}^n(\sigma(\Phi_{ij})\textbf{F}_{2i}-S_{ij}\textbf{F}_{2i})
 +2\beta_2(\textbf{G}_{2j}-\textbf{B}_{2j}) +2\mu_2(\textbf{G}_2-\textbf{W}\textbf{L})+2\nu_2\textbf{G}_2\textbf{1}
\end{split}
\end{equation}
Then we can compute the $\frac{\partial \mathcal{J}_2}{\partial \theta_{y_2}}$ according to the BP updating rule until convergency.

\textbf{Step 3}. Update $\textbf{B}$.

For the I2T retrieval task, the problem in Eq.(\ref{eq:W}) can be formulated as
\begin{equation}
\label{eq:11}
\begin{split}
\min \limits_{\textbf{B}_1}\,\lambda_1 {\lVert \textbf{B}_1 -\textbf{F}_1 \rVert}_F^2+\beta_1 {\lVert \textbf{B}_1-\textbf{G}_1 \rVert}_F^2  \quad s.t.\,\textbf{B}_1 \in{\left \{-1,1 \right \}}^{r\times n}
\end{split}
\end{equation}
The solution of Eq.(\ref{eq:11}) can be easily obtained by optimizing without relaxing discrete binary constraints $\textbf{B}_1\in{\left \{-1,1 \right\}}^{r\times n}$. Thus, we have
\begin{equation}
\label{eq:12}
\begin{split}
\textbf{B}_1=\texttt{sgn}(\lambda_1 \textbf{F}_1+\beta_1 \textbf{G}_1)
\end{split}
\end{equation}

Similar to the optimization of I2T retrieval task, the hash codes of T2I retrieval task can be obtained as
\begin{equation}
\label{eq:hhh}
\begin{split}
\textbf{B}_2=\texttt{sgn}(\lambda_2 \textbf{F}_2+\beta_2 \textbf{G}_2)
\end{split}
\end{equation}

\textbf{Step 4}. Update $\textbf{P}$, $\textbf{W}$.

For the I2T retrieval task, the optimization problem for updating $\textbf{P}$ can be simplified to
\begin{equation}
\label{eq:13}
\begin{split}
\min  \limits_{\textbf{P }}\,\mu_1{\lVert \textbf{F}_1 -\textbf{P}\textbf{L} \rVert}_F^2 +\nu_1{\lVert \textbf{P} \rVert}_F^2
\end{split}
\end{equation}

Letting the derivative of Eq.(\ref{eq:13}) with respect to $\textbf{P}$ to be equal to zero, we obtain
\begin{equation}
\label{eq:14}
\begin{split}
\textbf{P}=\textbf{F}_1\textbf{L}^\texttt{T}(\textbf{L}\textbf{L}^\texttt{T}+\frac{\nu_1}{\mu_1}\textbf{I})^{-1}
\end{split}
\end{equation}

For the T2I sub-retrieval task, the closed solution of $\textbf{W}$ can be calculated as
\begin{equation}
\label{eq:I}
\begin{split}
\textbf{W}=\textbf{G}_2\textbf{L}^\texttt{T}(\textbf{L}\textbf{L}^\texttt{T}+\frac{\nu_2}{\mu_2}\textbf{I})^{-1}
\end{split}
\end{equation}

The final results can be obtained by repeating the above steps until convergence. Algorithm \ref{alg:conjugateGradient} summarizes the key optimization steps for the I2T task in the proposed TA-ADCMH.

\begin{algorithm}[t]
\caption{\textbf{Discrete optimization for I2T}}
\label{alg:conjugateGradient}
\begin{algorithmic}
\Require
The raw image matrix $\textbf{X}$, text feature matrix $\textbf{Y}$, pair-wise semantic matrix $\textbf{S}$, point-wise semantic label $\textbf{L}$, hash code length $r$, the parameters $\lambda_1$, $\beta_1$, $\mu_1$, $\nu_1$.
\Ensure
 Hash codes matrix $\textbf{B}_1$, deep network parameters $\theta_{x_1}$ and $\theta_{y_1}$.
\State Randomly initialize $\textbf{P}$, $\textbf{B}_1$, $\theta_{x_1}$, $\theta_{y_1}$.
\State Construct the mini-batch $N_1$ and $N_2$ from $\textbf{X}$ and $\textbf{Y}$ by random sampling, $N_1 = N_2$=128. Initialize the iteration number $T_1$ = $\lceil n/N_1\rceil$, $T_2$ = $\lceil n/N_2\rceil$
\Repeat\\
\quad \textbf{For } \emph{iter} =1,$\cdots$, $T_1$ \textbf{do} \\
\quad \quad \quad Calculate $\emph{f}_1(\textbf{X}_i;{\theta_{x_1}})$ by forward propagation. \\
\quad \quad \quad Calculate the derivative according to Eq.(\ref{eq:7})\\
\quad \quad \quad Update deep model parameters $\theta_{x_1}$ by using back propagation.\\
\quad \textbf{end for}\\
\quad \textbf{For } \emph{iter} =1,$\cdots$, $T_2$ \textbf{do}\\
\quad \quad \quad Calculate $\emph{g}_1(\textbf{Y}_j;{\theta_{y_1}})$ by forward propagation.\\
\quad \quad \quad Calculate the derivative according to Eq.(\ref{eq:C})\\
\quad \quad \quad Update deep model parameters $\theta_{y_1}$ by using back propagation.\\
\quad \textbf{end for}\\
\quad \quad Update hash codes $\textbf{B}_1$ according to Eq.(\ref{eq:12}).\\
\quad \quad Update semantic projection matrix $\textbf{P}$ according to Eq.(\ref{eq:14}).
     \Until{convergence}
\end{algorithmic}
\end{algorithm}
\subsection{Online query hashing}
 As we discussed earlier, TA-ADCMH is a deep asymmetric cross-modal hashing method. It learns task-adaptive hash functions for different retrieval tasks. Specifically, given a new query instance $\textbf{x}_q$ with image modality, we can obtain its hash codes for I2T retrieval task by using the following formula
\begin{equation}
\label{eq:aa}
b_q=h(\textbf{x}_{q})=\texttt{sgn}(f_1(\textbf{x}_{q};\theta_{\textbf{x}_1})) .
\end{equation}

Similarly, given a query instance with text modality $\textbf{y}_q$, we can obtain the corresponding hash codes for T2I retrieval task by
\begin{equation}
\label{eq:a}
b_q=h(\textbf{y}_q)=\texttt{sgn}(g_2(\textbf{y}_q;\theta_{\textbf{y}_2})) .
\end{equation}
\subsection{Computation complexity analysis}
At each iteration of discrete optimization for I2T,
it takes $O(rn^2)$ for updating $\theta_{x_1}$ or $\theta_{y_1}$,
$O(rn)$ for updating $\mathbf{B}_1$,
and $O({c^2}n + rcn)$ for updating $\mathbf{P}$.
Finally, the computation complexity of the discrete optimization process of I2T is $O(iter*n^2)$,
where \emph{iter} is the number of iterations and $iter\ll n$.
The computation complexity of T2I is the same as I2T.

\section{Experimental setting}
\subsection{Evaluation datasets}
In our experiments, we use three public cross-modal retrieval datasets: MIR Flickr \cite{huiskes2008mir}, NUS-WIDE \cite{chua2009nus} and IAPR TC-12 \cite{IAPRTC},
for performance evaluation.
MIR Flickr is offered by the LIACS Media Lab at Leiden University. The images of MIR Flickr are downloaded from the social photography site Flickr. So far, the corresponding paper has had over 1,000 Google Scholar citations and this dataset has been downloaded for 63 thousand from universities and companies worldwide.
NUS-WIDE is a real-world web image database and created by Lab for Media Search in National University of Singapore. To the best of our knowledge, it is the largest publicly available multi-modal dataset used for testing multimedia retrieval performance.
The dataset IAPR TC-12 is originally constructed by the Department of Information Studies, University of Sheffield. The image collection of the dataset consists of still natural images, such as animals, cities, landscapes and many other aspects of contemporary life, taken from locations around the world.

We choose these datasets in our experiments for the following reasons:

(1) Each image in the MIR Flickr, NUS-WIDE and IAPR TC-12 has corresponding text description. The image and its corresponding text descriptions constitute an image-text pair for our cross-modal retrieval task.

(2) Our method is a supervised hashing method. All images in the three datasets have accurate semantic labels, which can be exploited as the supervision and guide the hash function learning during the training process. Also, the semantic labels can be used as the ground truth and evaluate the performance of cross-modal hashing methods.

(3) To the best of our knowledge, NUSWIDE is the largest publicly available multi-modal dataset to evaluate the cross-modal retrieval performance. It is suitable for testing the performance of our method on large-scale dataset.

(4) These datasets have been widely used in recent cross-modal hashing methods, such as DCMH \cite{jiang2017deep}, UCH \cite{UCH2019}, DLFH \cite{jiang2019discrete}, NRDH \cite{NRDH2020}, UKD \cite{UKD2020}, and JDSH \cite{JDSH2020}.

(5) Three datasets are public and all available with free of charge.

The following illustrates the basic descriptions of the experimental datasets:

\textbf{MIR Flickr} includes $25,000$ pairs of image-text instances collected from Flickr website. Each image is manually annotated with one or multiple of the 24 unique labels. We select those images which have at least 20 texts for our experiment, and finally obtain 20,015 multi-modal data. We randomly select 2, 000 multi-modal data as the query set, and use the remaining 18,015 multi-modal data as the retrieval set. Within the retrieval set, we randomly select 5,000 multi-modal as the training set. We describe each text as a 1,386-dimensional BOW vector. To be fair, the input of shallow methods are $4,096$-dimensional CNN feature, and the input of deep methods are original image pixels.

\textbf{NUS-WIDE} includes $269,648$ instances with 81 semantic labels downloaded from Flickr website. Considering the imbalance of label distribution, we select the top 21 most common categories and ultimately obtain $195,834$ image-text pairs as our final dataset.
In our experiments, we choose 2,000 pairs of instances for query, $193,834$ pairs of instances for retrieval and $10,000$ pairs of instances for training. The text of each instance is expressed as a $1,000$-dimensional BOW vector. For the traditional methods, the image of each instance is described by a deep feature with $4,096$-dimension. For the deep methods, each image uses the original pixel directly as the input.

\textbf{IAPR TC-12} \cite{IAPRTC} includes 20,000 image-text pairs which are annotated with 255 labels.
We use the entire dataset for our experiments and randomly select 2,000 image-text pairs to form the query set.
We use the remaining 18,000 pairs as the retrieval set and randomly select 5,000 image-text pairs within them as the training set.
The text for each instance is represented as a 2,912-dimensional BOW vector.
For the baseline methods, the input of shallow methods are $4,096$-dimensional CNN features and the input of deep methods are original image pixels.
\begin{figure*}[t]
\centering
\subfigure[I2T ]{\includegraphics[scale=0.28]{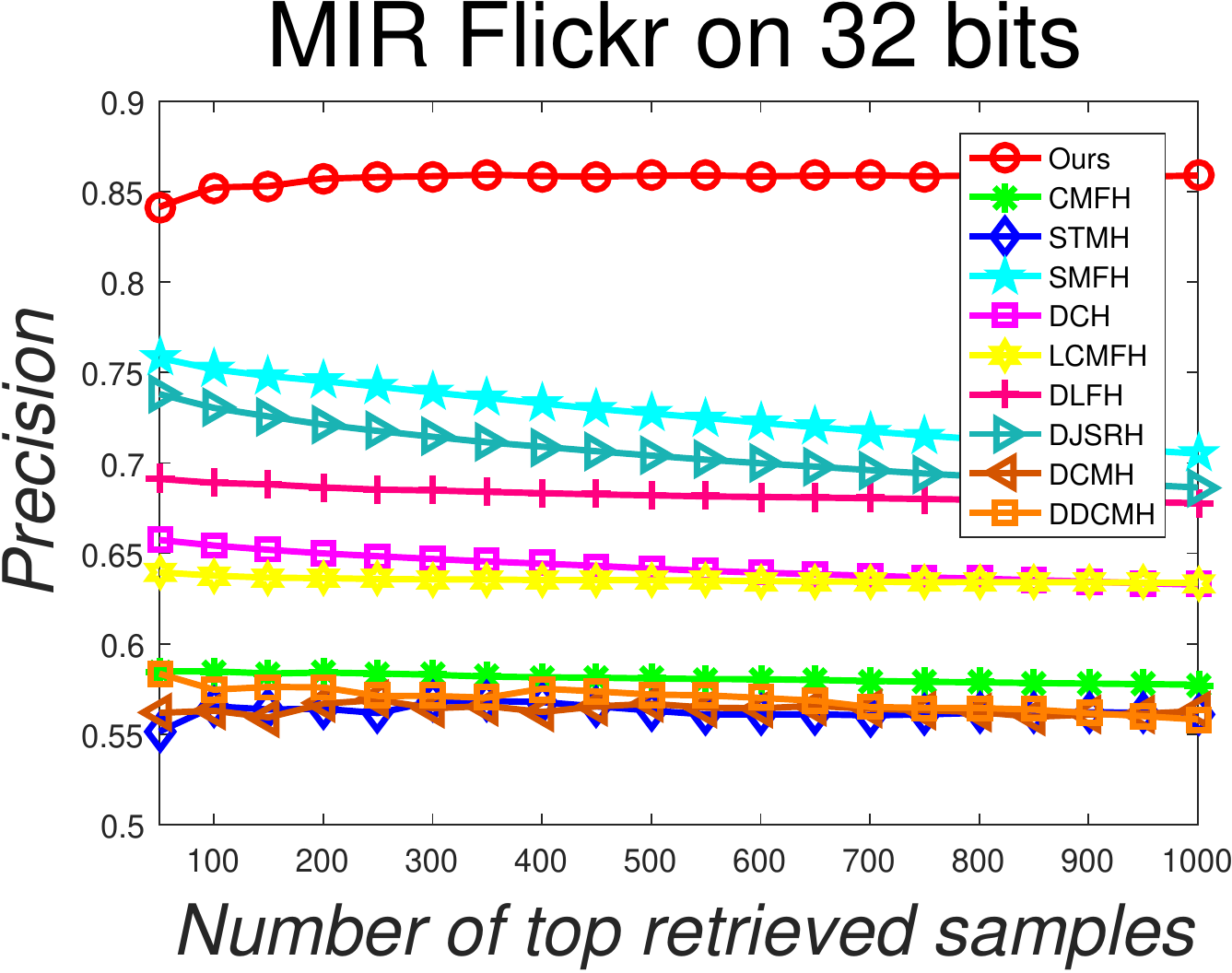}}
\mbox{}
\subfigure[I2T]{\includegraphics[scale=0.28]{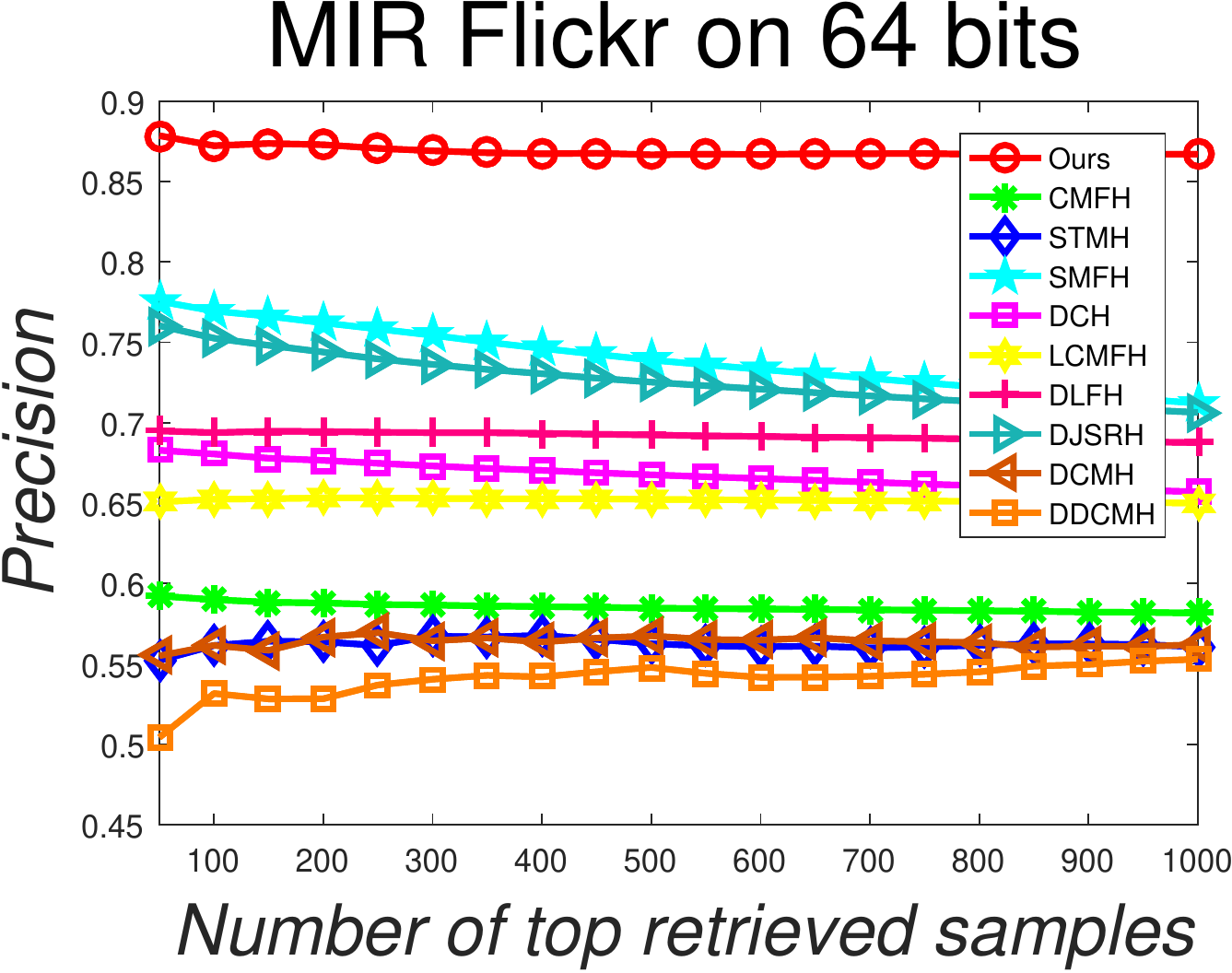}}
\mbox{}
\subfigure[I2T]{\includegraphics[scale=0.28]{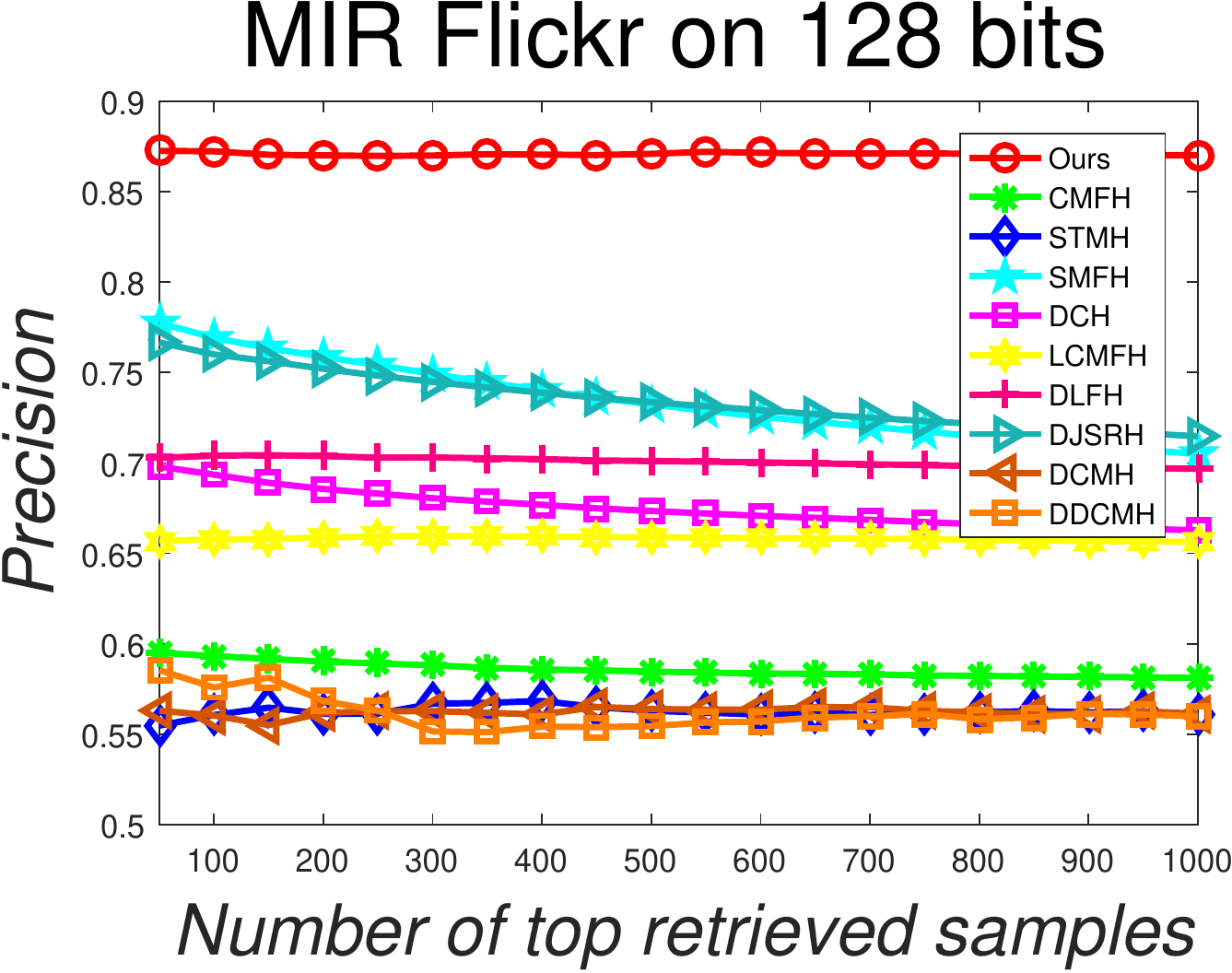}}
\\
\subfigure[T2I]{\includegraphics[scale=0.28]{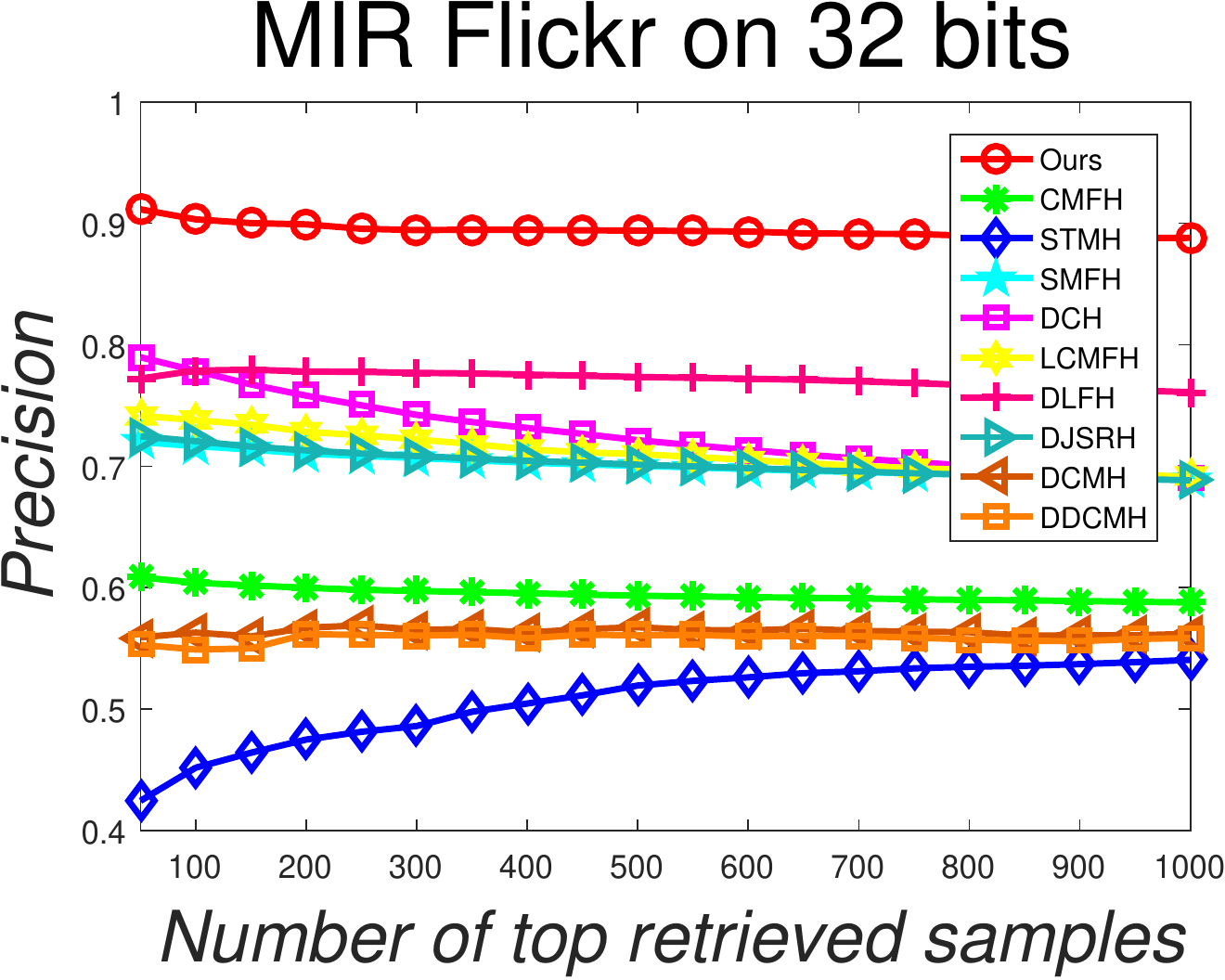}}
\mbox{}
\subfigure[T2I]{\includegraphics[scale=0.28]{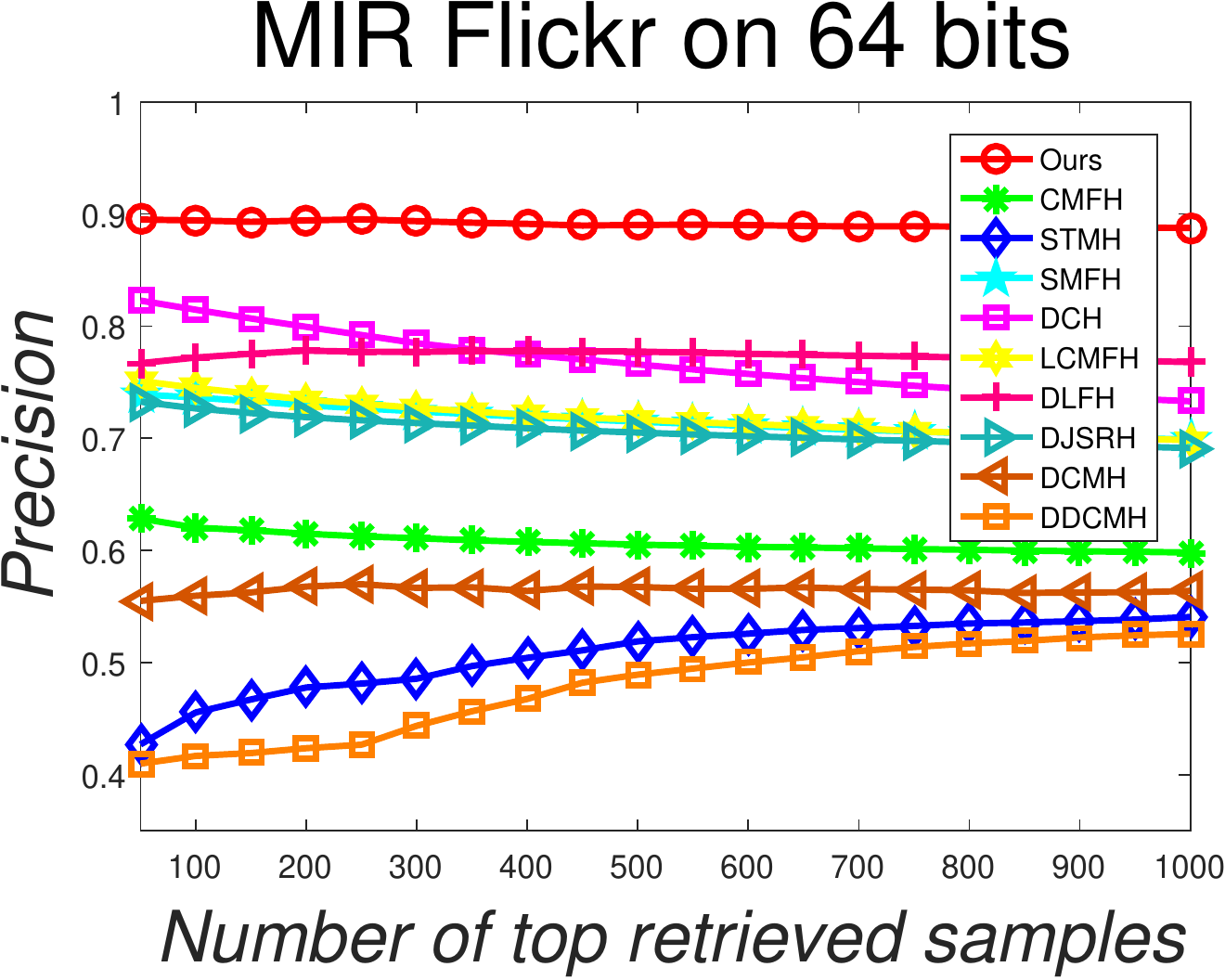}}
\mbox{}
\subfigure[T2I]{\includegraphics[scale=0.28]{T2I-top-mir-64.pdf}}
\vspace{-0.5cm}
\caption{The topK-precision results on MIR Flickr.}
\label{fig:2}
\end{figure*}
\subsection{Evaluation baselines and metrics}
We compare our proposed TA-ADCMH with several state-of-the-art cross-modal retrieval methods,
including two unsupervised methods: CMFH \cite{IEEEexample:cmfh}, DJSRH \cite{DJSRH2019},
and seven supervised methods: STMH \cite{wang2015semantic}, SMFH \cite{tang2016supervised}, DCH \cite{IEEEexample:ann6}, LCMFH \cite{wang2018label}, DLFH \cite{jiang2019discrete}, DCMH \cite{jiang2017deep}, DDCMH \cite{zhong2018deep}.
Among these methods, DJSRH, DCMH and DDCMH are deep methods while the others are shallow methods.
Note that there are two versions of DLFH that have different optimization algorithms. In our experiment, the kernel logistic regression is used for DLFH method.
To evaluate the retrieval performance, we adopt mean Average Precision (mAP) \cite{xu2016learning} and topK-precision \cite{lu2019efficient, Xu:2016:DLB:2964284.2967206} as the evaluation metrics.

%
%
\begin{table}[t]
\centering
\caption{{Retrieval performance comparison (mAP) on MIR Flickr.} }
\vspace{-0.4cm}
\setlength{\tabcolsep}{1.2mm}{
\begin{tabular}{|c|c|c|c|c|c|c|c|c|}
\hline
\multirow{2}{*}{Methods} & \multicolumn{4}{c|}{I2T } & \multicolumn{4}{c|}{T2I } \\ \cline{2-9}
 & 16 bits & 32 bits & 64 bits & 128 bits & 16 bits & 32 bits & 64 bits &128 bits \\ \hline
CMFH	&	0.6082 	&	0.6392 	&	0.6486 	&	0.6560 	&	0.6227 	&	0.6755 	&	0.7010 	&	0.7143 	\\\hline
STMH	&	0.7653 	&	0.7918 	&	0.8127 	&	0.8210 	&	0.7147 	&	0.7619 	&	0.7828 	&	0.8020 	\\\hline
SMFH	&	0.6132 	&	0.6173 	&	0.6185 	&	0.6157 	&	0.4820 	&	0.4867 	&	0.4881 	&	0.4876 	\\\hline
DCH	    &	0.6945 	&	0.7124 	&	0.7327 	&	0.7433 	&	0.8169 	&	0.8235 	&	0.8564 	&	0.8591 	\\\hline
LCMFH	&	0.6747 	&	0.6846 	&	0.6979 	&	0.7061 	&	0.7471 	&	0.7780 	&	0.7874 	&	0.7843 	\\\hline
DLFH	&	0.7063 	&	0.7356 	&	0.7346 	&	0.7382 	&	0.7600 	&	0.7951 	&	0.7897 	&	0.8043 	\\\hline
DJSRH	&	0.7623 	&	0.7853 	&	0.8110 	&	0.8115 	&	0.6947 	&	0.7683 	&	0.7797 	&	0.7869 	\\\hline
DCMH	&	0.6076 	&	0.6081 	&	0.6061 	&	0.6081 	&	0.6063 	&	0.6100 	&	0.6066 	&	0.6280 	\\\hline
DDCMH	&	0.6164 	&	0.6184 	&	0.6048 	&	0.6532 	&	0.5959 	&	0.5974 	&	0.5037 	&	0.5968 	\\\hline
Ours	&	0.8622 	&	0.8758 	&	0.8839 	&	0.8884 	&	0.9053 	&	0.9083 	&	0.9060 	&	0.9159 	\\\hline
\end{tabular}}\label{lab:3}
\end{table}
\begin{figure}[t]
\centering
\subfigure[I2T]{\includegraphics[scale=0.28]{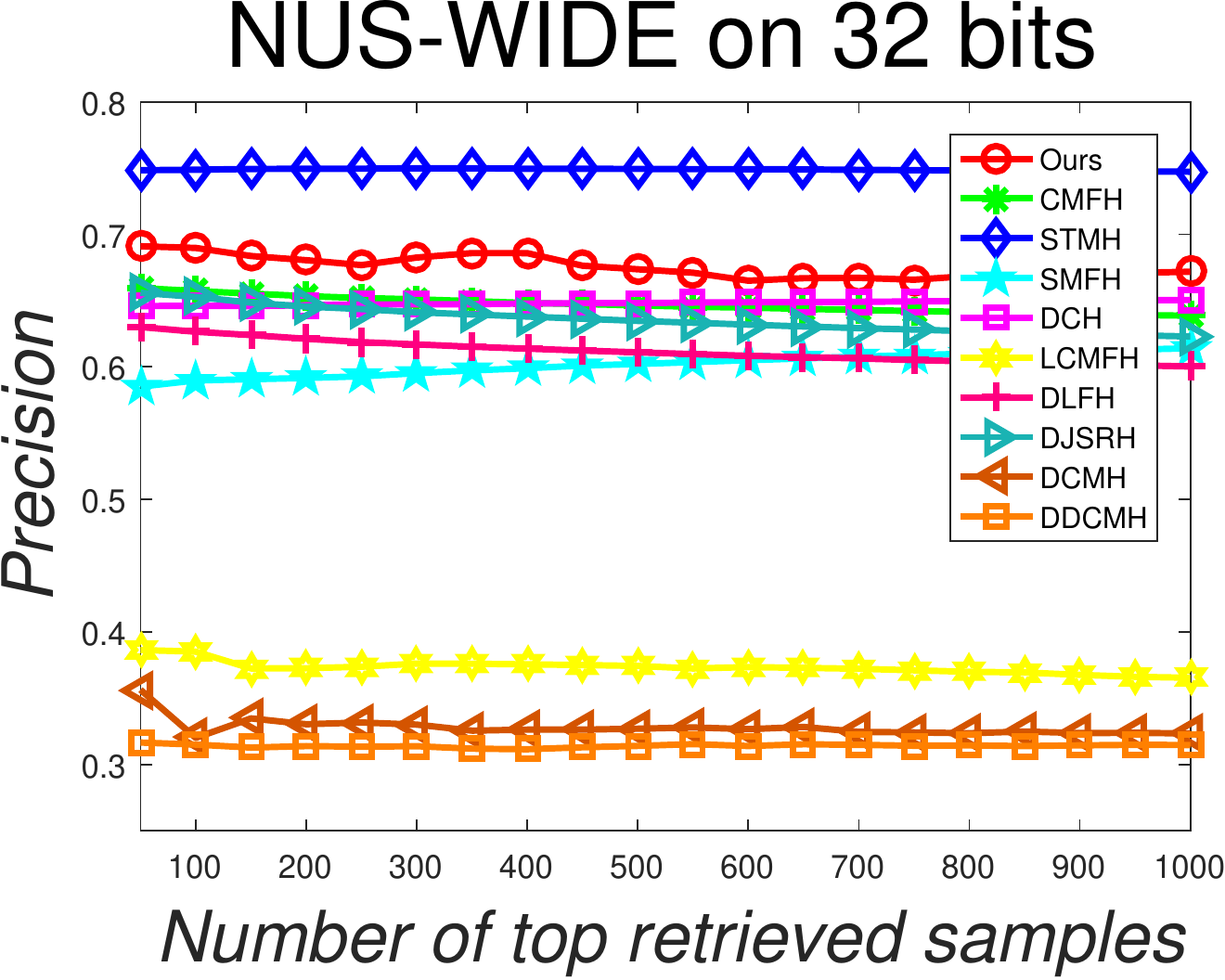}}
\mbox{}
\subfigure[I2T]{\includegraphics[scale=0.28]{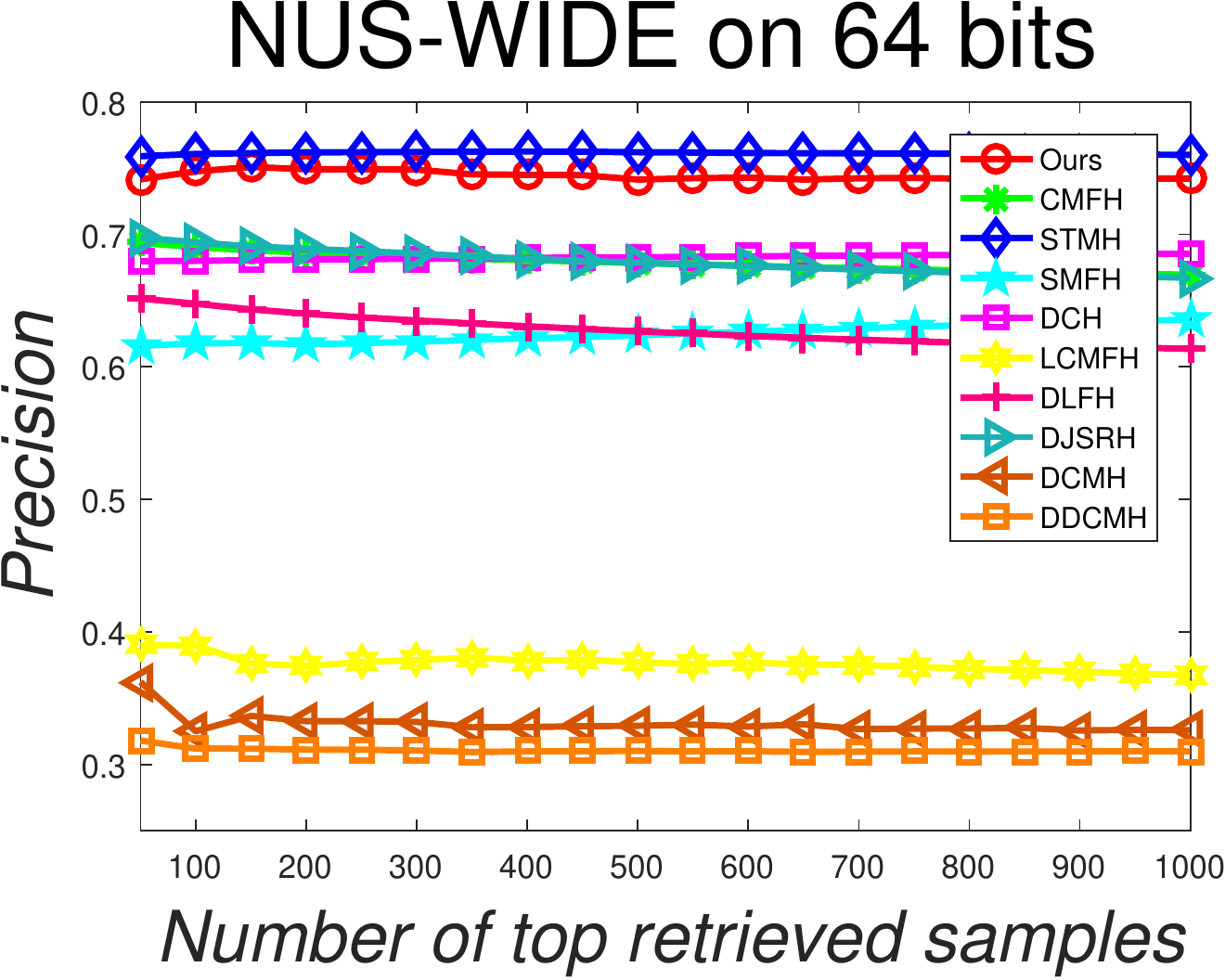}}
\mbox{}
\subfigure[I2T]{\includegraphics[scale=0.28]{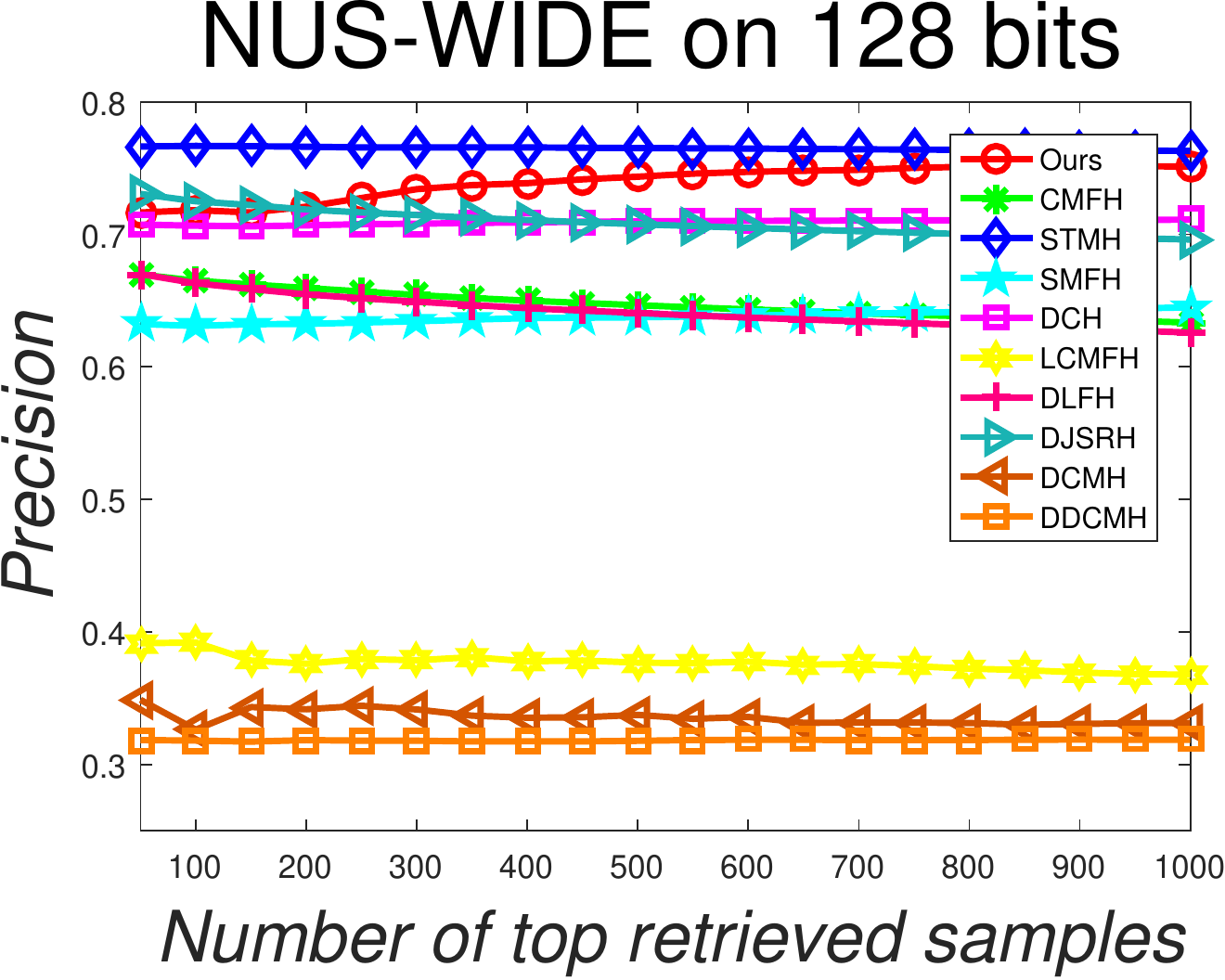}}
\\
\subfigure[T2I]{\includegraphics[scale=0.28]{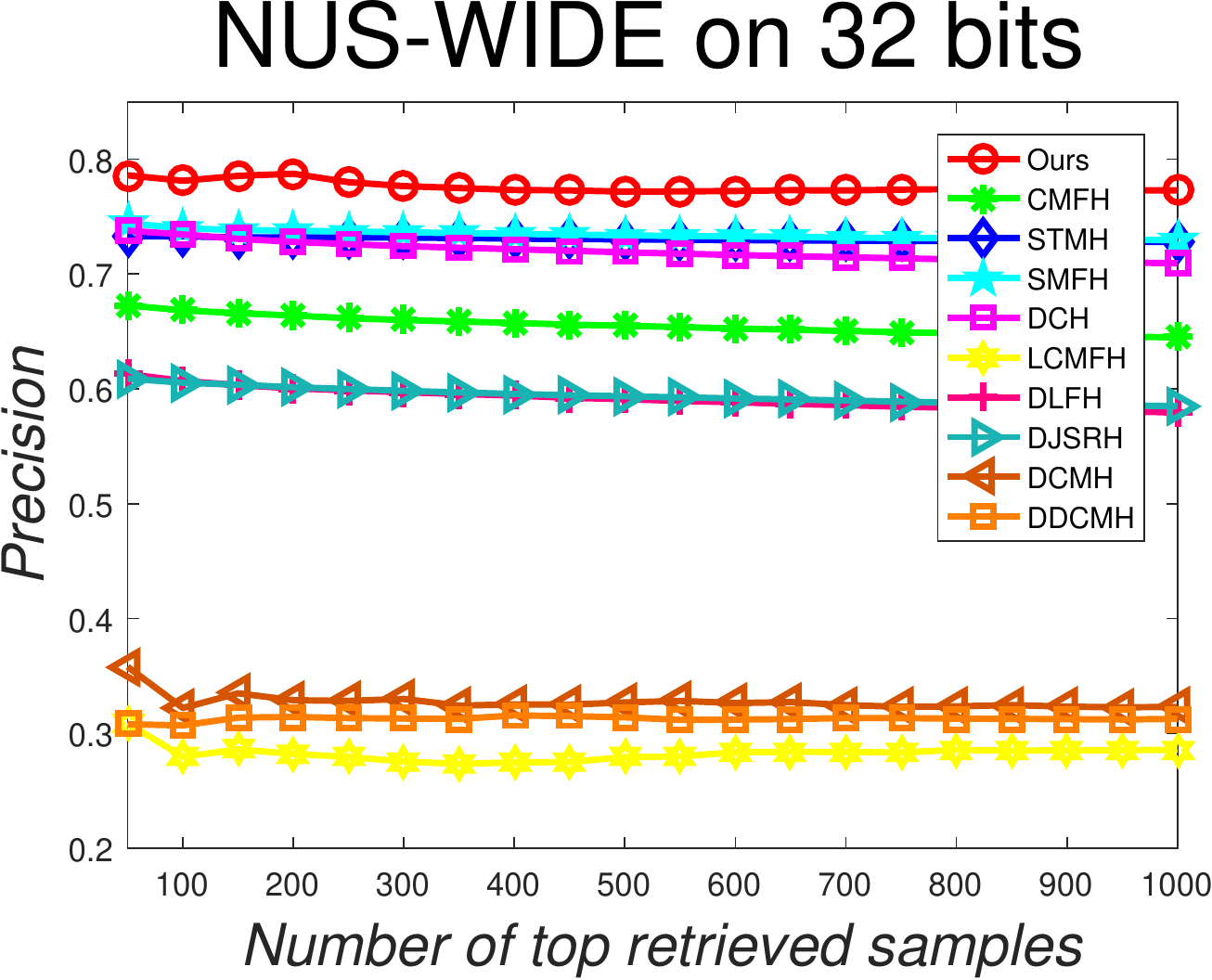}}
\mbox{}
\subfigure[T2I]{\includegraphics[scale=0.28]{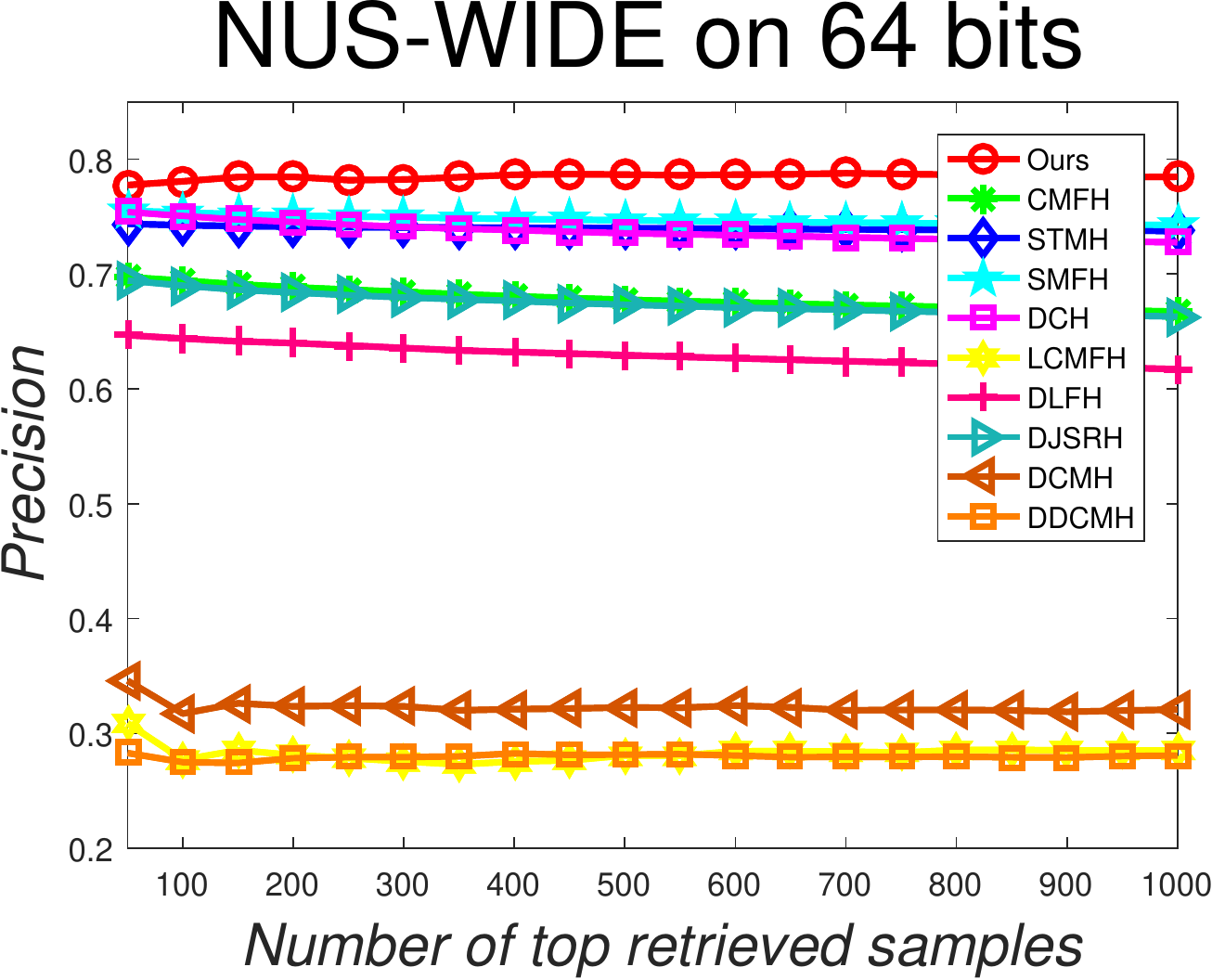}}
\mbox{}
\subfigure[T2I]{\includegraphics[scale=0.28]{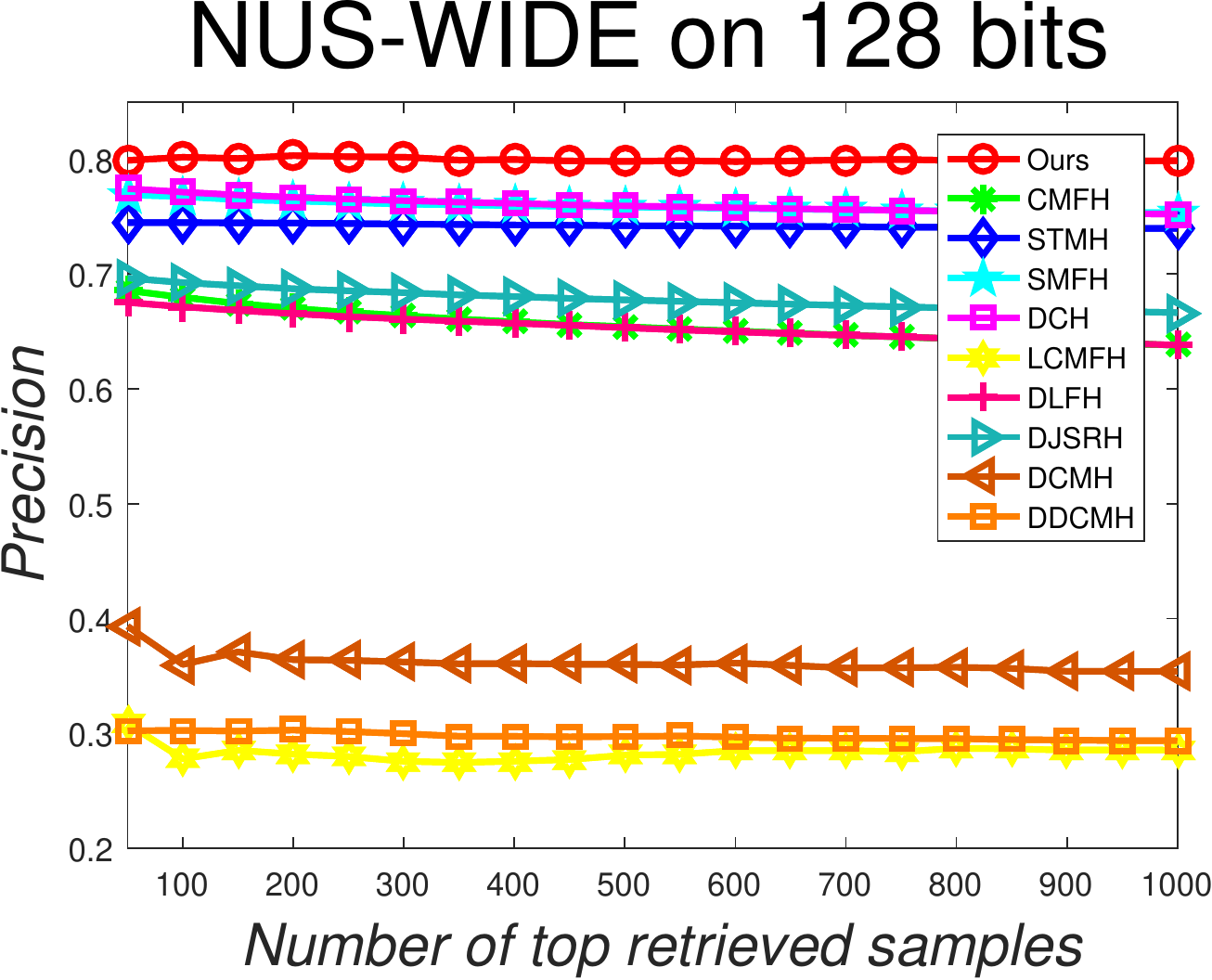}}
\vspace{-0.6cm}
\caption{The topK-precision results on NUS-WIDE.}
\label{fig:3}
\end{figure}
\subsection{Implementation details}
Our method formulates two objective functions, which consist of eight parameters: $\lambda_1$, $\beta_1$, $\mu_1$, $\nu_1$, $\lambda_2$, $\beta_2$, $\mu_2$, $\nu_2$ for two different cross-modal retrieval tasks. In the task of I2T retrieval, the regularization parameters $\lambda_1$, $\beta_1$ control the regression from the deep features to the asymmetric binary codes for images and text respectively. The parameter $\mu_1$ ensures the deep features of image modality are ideal for  supervised classification, $\nu_1$ is the regularization parameter to avoid overfitting. The parameters of the T2I retrieval task are similar to I2T task. We set different values for the involved parameters to optimize the I2T and T2I retrieval tasks. In our experiment, the best performance of I2T task can be achieved when $\{\lambda_1=10^{-1}, \beta_1=10^{-2}, \mu_1=10^{-4}, \nu_1=10^{-1}\}$, and that of T2I task can be achieved when $\{\lambda_2=10^{-1}, \beta_2=10^{-2}, \mu_2=10^{-4}, \nu_2=10^{-1}\}$ on MIR Flickr.
The best performance can be obtained for I2T task when $\{\lambda_1=10^{-1}, \beta_1=1, \mu_1=1, \nu_1=10^{-1}\}$, and that for T2I task when $\{\lambda_2=10^{-1}, \beta_2=1, \mu_2=1, \nu_2=10^{-1}\}$ on NUS-WIDE.
The best performance can be obtained for I2T task when $\{\lambda_1=1, \beta_1=10^{-5}, \mu_1=0.1, \nu_1=0.1\}$, and that for T2I task when $\{\lambda_2=1, \beta_2=10, \mu_2=0.1, \nu_2=10^{-5}\}$ on IAPR TC-12.
In all cases, the number of iterations for the outer-loop is set to 500. Moreover, we implement our method on Matconvnet and use CNN deep networks which are the same as DCMH. Before the training process, we initialize the weights after pretreatment of the original data on the ImageNet dataset. In the network learning process, we use the raw pixels for images and the BOW vectors for texts as inputs to the deep networks, respectively. The learning rate is in the range of $[10^{-6}, 10^{-1}]$. The batch size is set to 128 for two couples of deep networks. All of the results are conducted 3 times and the average results are presented. All experiments are carried out on a computer with Intel(R) Xeon(R) Gold 6130 CPU @ 2.10GHz and 64-bit Ubuntu 16.04.6 LTS operating system.

\section{Experimental results}
\subsection{Performance comparison}
In the experiments, we first report the mAP values of the compared methods on three datasets. The mAP results of all baselines with the hash code length ranging from 16 bits to 128 bits are presented in Table \ref{lab:3} - Table \ref{tabcoco1}. On the basis of these results, we can reach the following conclusion:
1) For both I2T and T2I sub-retrieval tasks, TA-ADCMH outperforms the compared methods with different codes lengths in most cases. These results clearly prove the feasibility of our method. The main reasons for the superior performance are: Firstly, TA-ADCMH trains two couples of deep neural networks to perform different retrieval tasks independently, which can enhance the nonlinear representation of deep features and capture the query semantics of two sub-retrieval tasks. Secondly, we jointly adopt the pair-wise and point-wise semantic labels to generate binary codes which can express semantic similarity of different modalities.
2) It is noteworthy that the mAP values of TA-ADCMH is in the upward trend with the increase of the code length.
These results demonstrate that longer binary codes  have stronger discriminative capability with effective discrete optimization in our method.
3) The results of most methods have higher mAP values on the T2I task than that obtained on I2T task. This depends on the fact that text features can better reflect the semantic information of instances.
4) The deep TA-ADCMH method makes significant improvement compared with the shallow methods on three datasets.
This phenomenon is attributed to the deep feature representation extracted by the nonlinear projections and the semantic information used in binary hash mapping.
The results show that the deep neural network has better representation capability.
5) We can find that our approach, TA-ADCMH, achieves higher performance than the unsupervised methods over the T2I task and I2T task on three datasets. This is due to the fact that our method can learn more discriminative hash codes through the supervision of explicit semantic labels.

\begin{table}[t]
\centering
\caption{{Retrieval performance comparison (mAP) on NUS-WIDE.} }
\setlength{\tabcolsep}{1.2mm}{
\begin{tabular}{|c|c|c|c|c|c|c|c|c|}
\hline
\multirow{2}{*}{Methods} & \multicolumn{4}{c|}{I2T} & \multicolumn{4}{c|}{T2I} \\ \cline{2-9}
 & 16 bits & 32 bits & 64 bits & 128 bits & 16 bits & 32 bits & 64 bits &128 bits \\ \hline
CMFH	&	0.6671 	&	0.7083 	&	0.7353 	&	0.7260 	&	0.6884 	&	0.7154 	&	0.7404 	&	0.7383 	\\\hline
STMH	&	0.6373 	&	0.7179 	&	0.7715 	&	0.7842 	&	0.6203 	&	0.6519 	&	0.6997 	&	0.7170 	\\\hline
SMFH	&	0.5221 	&	0.5240 	&	0.5282 	&	0.5280 	&	0.3568 	&	0.3606 	&	0.3691 	&	0.3653 	\\\hline
DCH	&	0.6524 	&	0.6917 	&	0.7207 	&	0.7431 	&	0.7581 	&	0.7699 	&	0.7826 	&	0.7976 	\\\hline
LCMFH	&	0.5235 	&	0.5543 	&	0.5868 	&	0.5983 	&	0.4830 	&	0.4955 	&	0.5019 	&	0.4826 	\\\hline
DLFH	&	0.6030 	&	0.6356 	&	0.6737 	&	0.6960 	&	0.7137 	&	0.7807 	&	0.7860 	&	0.8022 	\\\hline
DJSRH	&	0.6408 	&	0.7035 	&	0.7400 	&	0.7702 	&	0.5194 	&	0.6594 	&	0.7447 	&	0.7430 	\\\hline
DCMH	&	0.4302 	&	0.4304 	&	0.4325 	&	0.4394 	&	0.4302 	&	0.4310 	&	0.4233 	&	0.4687 	\\\hline
DDCMH	&	0.4367 	&	0.4234 	&	0.4186 	&	0.4087 	&	0.3573 	&	0.4088 	&	0.3829 	&	0.3968 	\\\hline									Ours	&	0.7552 	&	0.7892 	&	0.7913 	&	0.7939 	&	0.7968 	&	0.8001 	&	0.8127 	&	0.8257 	\\\hline
\end{tabular}}\label{lab:4}
\end{table}
\begin{figure}[t]
\centering
\subfigure[I2T]{\includegraphics[scale=0.28]{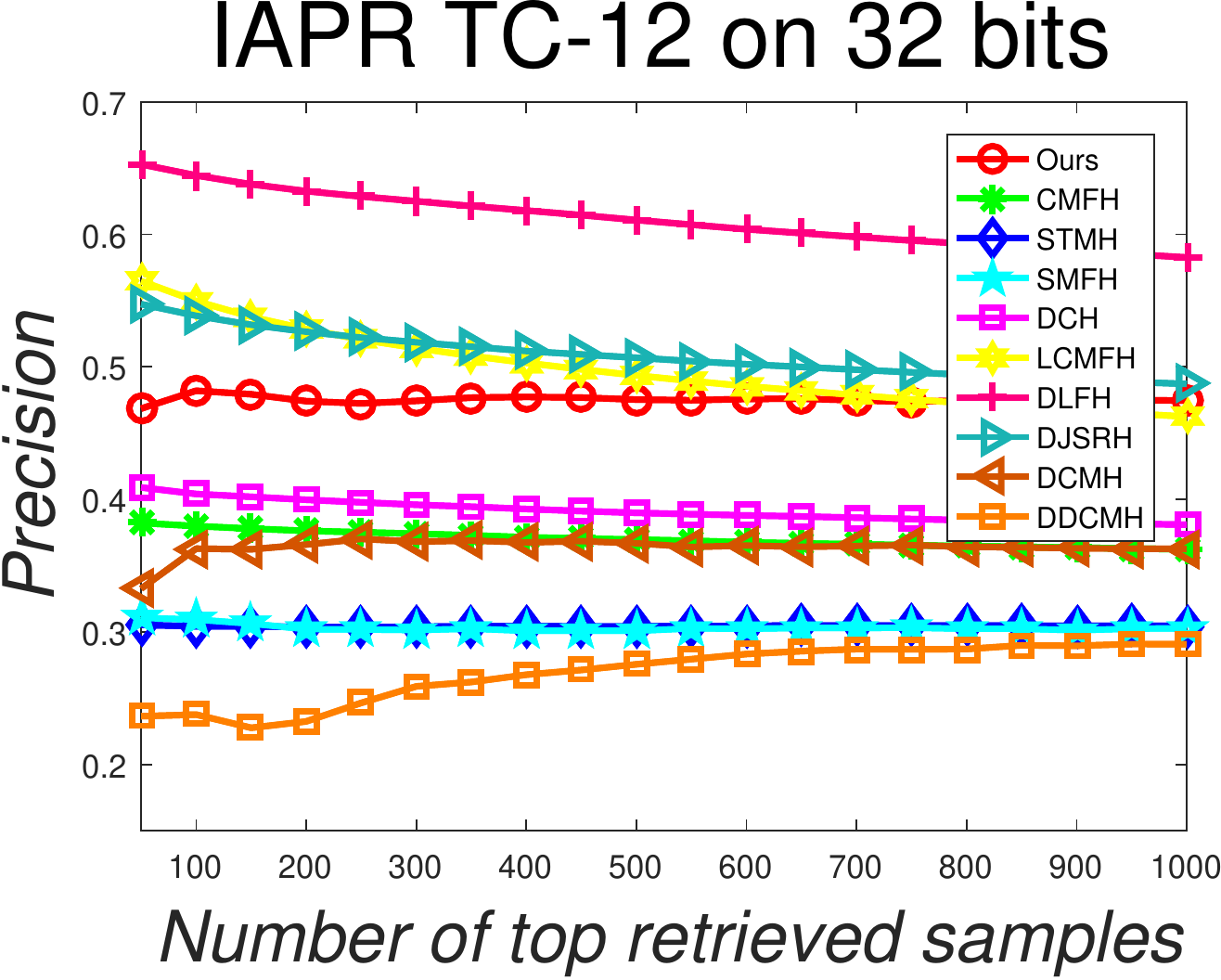}}
\mbox{}
\subfigure[I2T]{\includegraphics[scale=0.28]{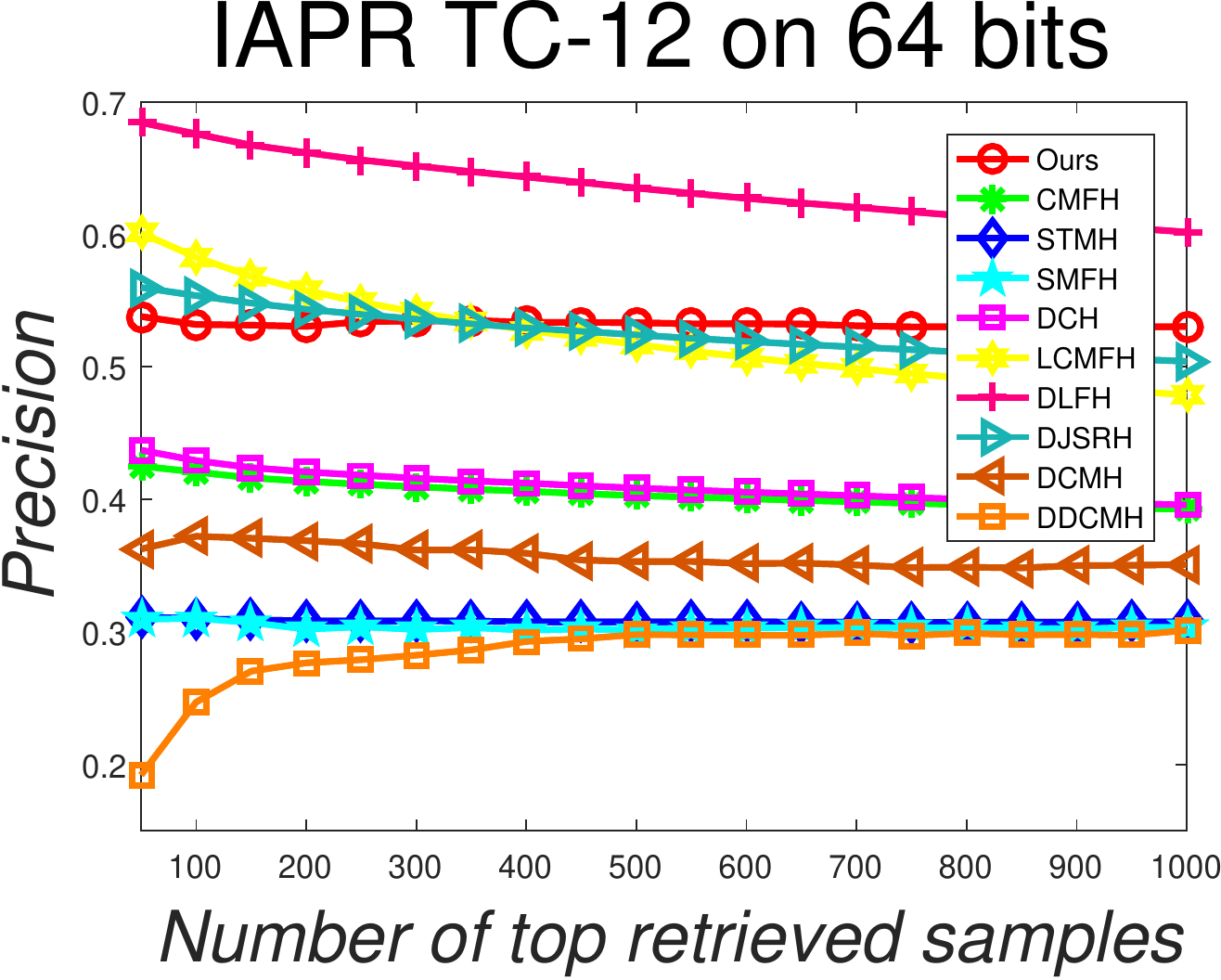}}
\mbox{}
\subfigure[I2T]{\includegraphics[scale=0.28]{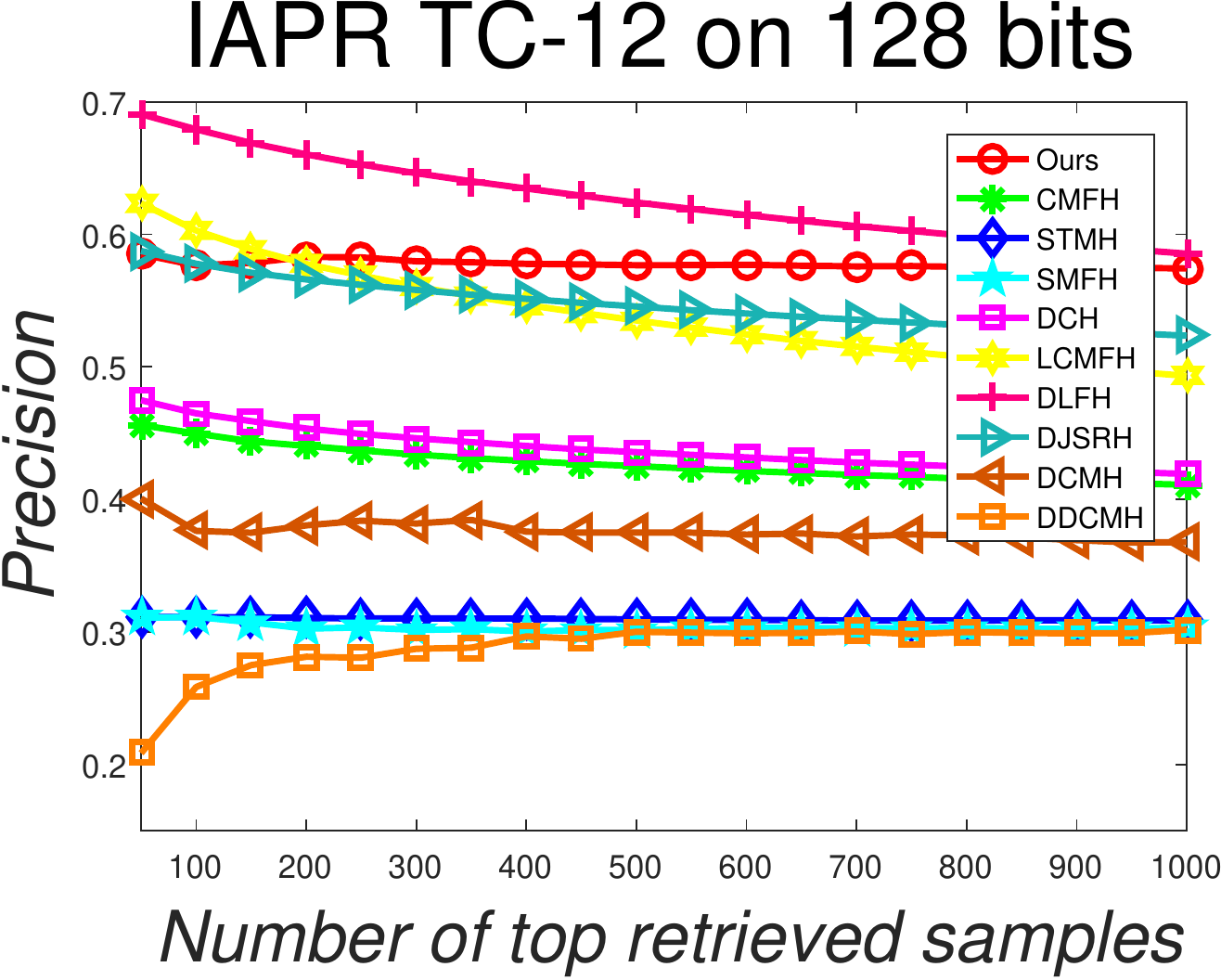}}
\\
\subfigure[T2I]{\includegraphics[scale=0.28]{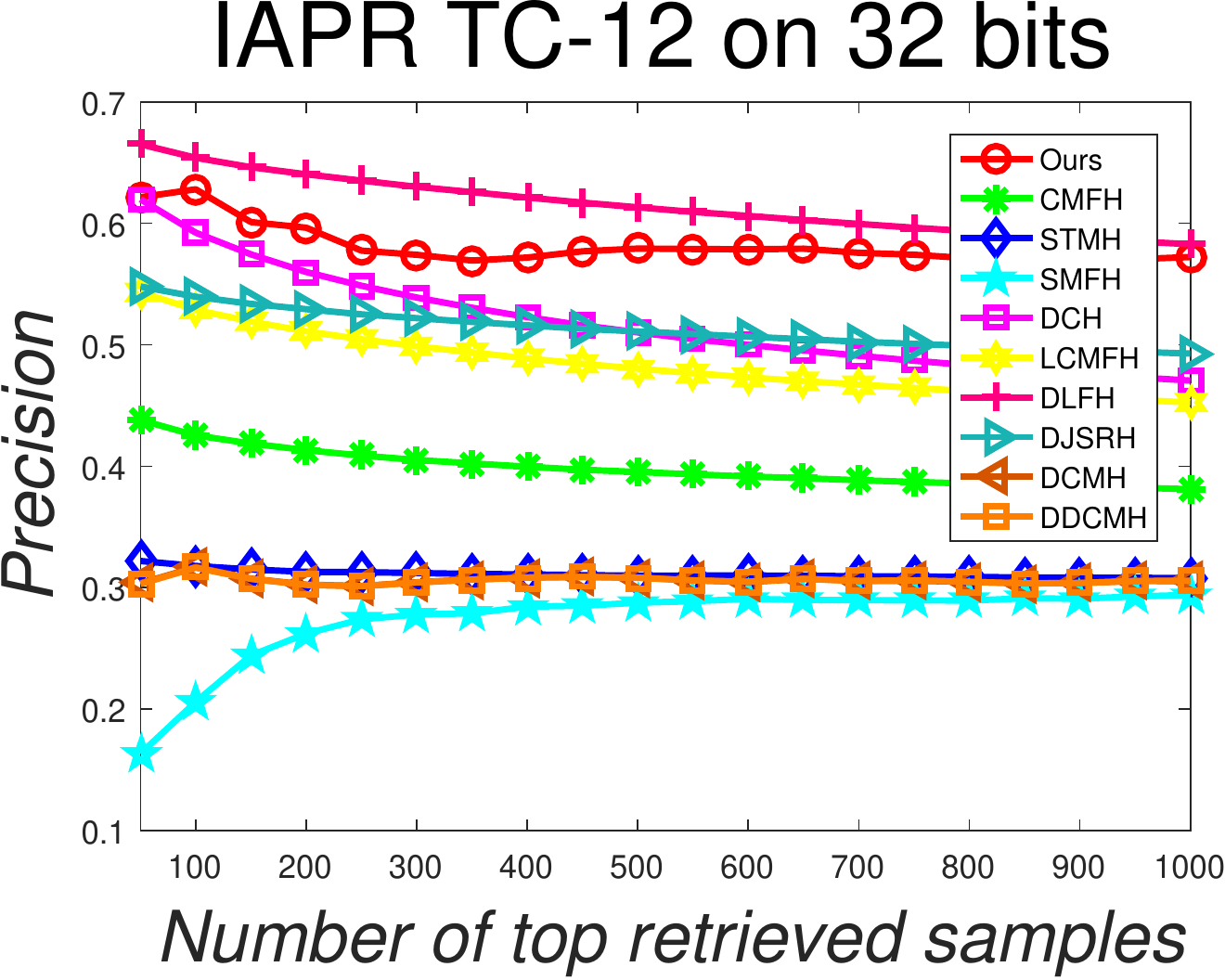}}
\mbox{}
\subfigure[T2I]{\includegraphics[scale=0.28]{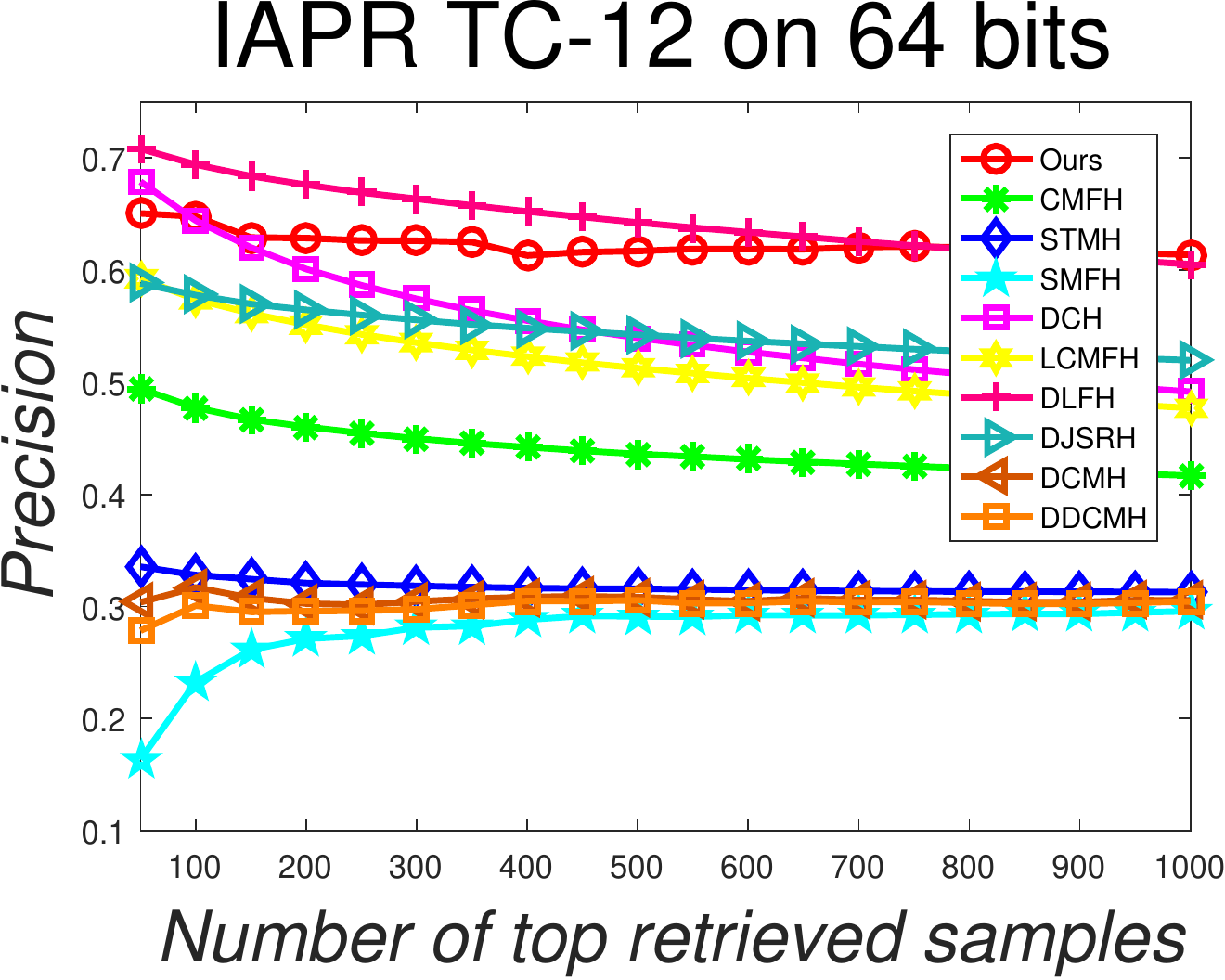}}
\mbox{}
\subfigure[T2I]{\includegraphics[scale=0.28]{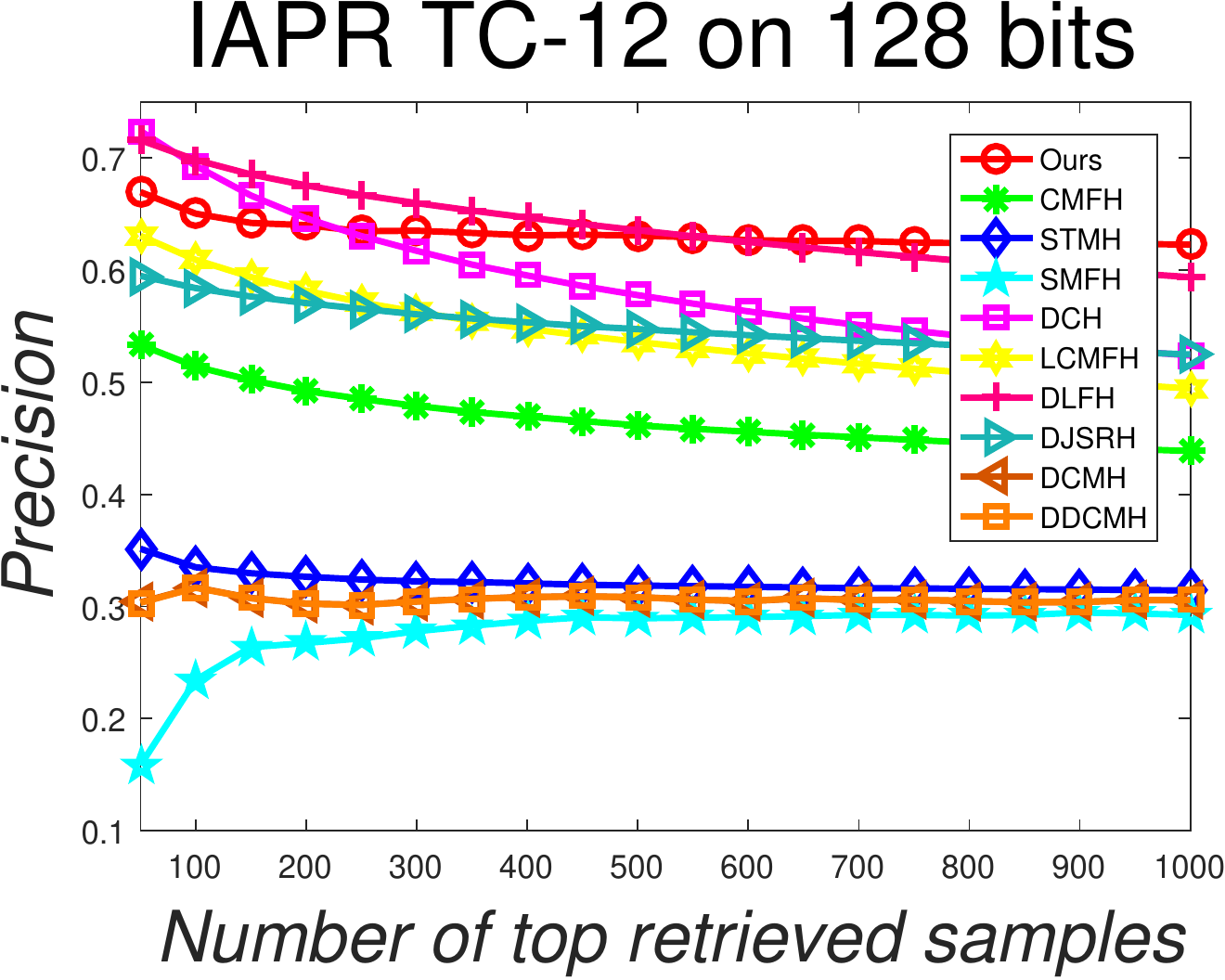}}
\vspace{-0.6cm}
\caption{The topK-precision results on IAPR TC-12.}
\label{figcoco1}
\end{figure}
\begin{table}[t]
\centering
\caption{{Retrieval performance comparison (mAP) on IAPR TC-12.} }
\setlength{\tabcolsep}{1.2mm}{
\begin{tabular}{|c|c|c|c|c|c|c|c|c|}
\hline
\multirow{2}{*}{Methods} & \multicolumn{4}{c|}{I2T } & \multicolumn{4}{c|}{T2I } \\ \cline{2-9}
 & 16 bits & 32 bits & 64 bits & 128 bits & 16 bits & 32 bits & 64 bits &128 bits \\ \hline
CMFH	&	0.4476 	&	0.4658 	&	0.5092 	&	0.5363 	&	0.4837 	&	0.5465 	&	0.6081 	&	0.6491 	\\\hline
STMH	&	0.5143 	&	0.5516 	&	0.5897 	&	0.6379 	&	0.4969 	&	0.5612 	&	0.5984 	&	0.6665 	\\\hline
SMFH	&	0.4170 	&	0.4105 	&	0.4096 	&	0.4104 	&	0.2866 	&	0.2525 	&	0.2396 	&	0.2320 	\\\hline
DCH	&	0.4822 	&	0.4918 	&	0.5160 	&	0.5520 	&	0.6658 	&	0.7115 	&	0.7681 	&	0.8022 	\\\hline
LCMFH	&	0.4327 	&	0.4390 	&	0.4463 	&	0.4733 	&	0.4661 	&	0.4940 	&	0.5533 	&	0.5940 	\\\hline
DLFH	&	0.4593 	&	0.4996 	&	0.5208 	&	0.5372 	&	0.5112 	&	0.5590 	&	0.6040 	&	0.6278 	\\\hline
DJSRH	&	0.6036 	&	0.6200 	&	0.6310 	&	0.6520 	&	0.5861 	&	0.6160 	&	0.6560 	&	0.6610 	\\\hline
DCMH	&	0.4538 	&	0.4357 	&	0.4208 	&	0.4411 	&	0.3765 	&	0.3765 	&	0.3765 	&	0.3765 	\\\hline
DDCMH	&	0.4222 	&	0.3365 	&	0.2339 	&	0.2645 	&	0.2412 	&	0.4243 	&	0.2640 	&	0.4247 	\\\hline
Ours	&	0.6139 	&	0.6391 	&	0.6660 	&	0.7151 	&	0.6659 	&	0.6287 	&	0.6676 	&	0.6756 	\\\hline
\end{tabular}}\label{tabcoco1}
\end{table}
Next, we illustrate the performance comparison in terms of the topK precision from 32 bits to 128 bits on the three datasets.
Figure \ref{fig:2} - Figure \ref{figcoco1} illustrate the results on MIR Flickr, NUS-WIDE and IAPR TC-12, respectively.
As shown by these figures, TA-ADCMH can obtain the higher precision and reliability compared with other baselines in most cases
on both I2T and T2I sub-retrieval tasks with different code lengths.
Moreover, we can also observe that the topK-precision curves of TA-ADCMH is relatively stable with increasing numbers of retrieved samples $K$. These observations are enough to prove that TA-ADCMH has strong capability to retrieve the relevant samples effectively. Compared with I2T retrieval task, the topK-precision of TA-ADCMH can obtain much better performance than the baseline methods on T2I cross-modal retrieval task. It is consistent with the mAP values in Table \ref{lab:3} - Table \ref{tabcoco1}. In practice, users browse the website according to the ranking of retrieval results, so they are interested on the top-ranked instances in the retrieved list. Thus, TA-ADCMH significantly outperforms the compared methods on two sub-retrieval tasks.

To summarize, TA-ADCMH achieves superior performance on MIR Flickr, NUS-WIDE and IAPR TC-12. This phenomenon validates that capturing the query semantics of different cross-modal retrieval tasks is effective when learning the cross-modal hash codes. All the results confirm the effectiveness of our designed loss functions and optimization scheme.

In addition, we use the statistical approach of t-test on three datasets to analyze whether
there are meaningful differences between our generated hash codes
and several competitive baseline methods by observing the p-values shown in Table \ref{ttest}.
From the experimental results, we can see that our method not only has superior mAP performance,
but also represents statistically significant improvements over the comparison methods in most cases.
\begin{table}[t]
\small
\centering
\caption{T-test results on three datasets. 
h = 1 implies that t-test rejects the original hypothesis at the default significance level of 5\%.}
\setlength{\tabcolsep}{1mm}{
\begin{tabular}{|c|c|c|c|c|c|c|c|c|c|c|c|c|}
\hline
\multirow{3}{*}{Methods} & \multicolumn{4}{c|}{MIR Flickr} & \multicolumn{4}{c|}{NUS-WIDE} & \multicolumn{4}{c|}{IAPR TC-12}\\ \cline{2-13}
 & \multicolumn{2}{c|}{I2T} & \multicolumn{2}{c|}{T2I} & \multicolumn{2}{c|}{I2T} & \multicolumn{2}{c|}{T2I} & \multicolumn{2}{c|}{I2T} & \multicolumn{2}{c|}{T2I} \\ \cline{2-13}
 & h & p & h & p  & h & p & h & p & h & p & h & p \\ \hline
CMFH	&	1	&	1.87E-05	&	1	&	1.15E-03	&	1	&	1.95E-03	&	1	&	1.32E-03	&	1	&	4.89E-05	&	0	&	7.96E-02	\\\hline
STMH	&	1	&	1.27E-03	&	1	&	3.56E-03	&	0	&	1.17E-01	&	1	&	3.36E-03	&	1	&	5.68E-04	&	0	&	9.84E-02	\\\hline
DCH	&	1	&	1.59E-05	&	1	&	2.90E-07	&	1	&	6.47E-05	&	1	&	2.97E-05	&	1	&	1.70E-03	&	1	&	1.62E-04	\\\hline
SMFH	&	1	&	8.23E-05	&	1	&	5.64E-03	&	1	&	6.99E-03	&	1	&	8.84E-04	&	1	&	1.81E-04	&	0	&	6.62E-02	\\\hline
LCMFH	&	1	&	2.12E-06	&	1	&	5.22E-04	&	1	&	2.02E-04	&	1	&	4.29E-05	&	1	&	5.12E-04	&	1	&	1.30E-02	\\\hline
DLFH	&	1	&	2.28E-05	&	1	&	6.09E-04	&	1	&	2.43E-03	&	0	&	8.49E-02	&	1	&	3.60E-04	&	1	&	3.97E-02	\\\hline
DJSRH	&	1	&	8.46E-04	&	1	&	4.73E-03	&	1	&	4.03E-02	&	0	&	5.86E-02	&	0	&	7.08E-02	&	0	&	1.74E-01	\\\hline
DCMH	&	1	&	2.23E-05	&	1	&	1.78E-06	&	1	&	2.80E-05	&	1	&	1.37E-05	&	1	&	3.01E-03	&	1	&	1.11E-04	\\\hline
DDCMH	&	1	&	1.11E-04	&	1	&	6.46E-04	&	1	&	1.45E-04	&	1	&	3.51E-05	&	1	&	1.08E-02	&	1	&	9.95E-03	\\\hline
\end{tabular}}\label{ttest}
\end{table}
\begin{table}
\caption{Training time (Ttime) and Query time (Qtime) on three datasets with 32 bits. The results are reported in seconds.}
\centering
\begin{tabular}{|c|c|c|c|c|c|c|}
\hline
\multirow{2}{*}{Methods}& \multicolumn{2}{c|}{MIR Flickr}& \multicolumn{2}{c|}{NUS-WIDE}& \multicolumn{2}{c|}{IAPR TC-12}\\ \cline{2-7}
 & Ttime & Qtime & Ttime & Qtime & Ttime & Qtime\\
\hline			
CMFH	&	930.08	&35.86	&1044.03  &351.49	&1173.50 &36.60	\\\hline
STMH	&	71.20	&37.56	&167.74   &359.48	&94.34	 &39.45	\\\hline
DCH	    &	33.70	&35.65	&281.54   &344.99	&42.93	 &36.33	\\\hline
SMFH	&	42.69	&35.65	&122.24   &346.22	&31.65	 &36.36 \\\hline
LCMFH	&	103.54	&35.72	&296.14	  &344.97	&189.99	 &36.54	\\\hline
DLFH	&	12.01	&35.61	&25.89	  &345.53	&12.20	 &36.29 \\\hline
DJSRH	&	369.54	&17.93	&988.17	  &220.69	&373.59	 &17.38 \\\hline
DCMH	&	979.13	&36.18	&11390.70 &353.21	&5752.37 &36.18 \\\hline
DDCMH	&	17.81	&35.99	&36.78	  &347.36	&17.84	&36.82	\\\hline
TA-ADCMH&	1885.50	&	36.14 	&	23329.55 	&	353.28 	&	11492.17 	&	35.82 	\\\hline						
\end{tabular}
\label{RunTime}
\end{table}

Furthermore, we conduct experiments to compare the training and query time between our method and the baselines on three datasets when the code length is fixed to 32 bits.
The shallow methods use 4,096-dimensional CNN feature extracted by CNN-F in advance.
For fair comparison, we add the computation time consumed by feature extraction process to the training and query time of the shallow methods.
Therefore, the time of our method and the baseline methods are all calculated from the original images to image hash codes.
The comparison results are presented in Table \ref{RunTime}.
We can observe that the proposed TA-ADCMH consumes the most time at the training stage.
This is because our TA-ADCMH uses CNN-F model as the deep hash model and adpots an asymmetric learning strategy to learn two couples of deep hash functions for I2T retrieval task and T2I retrieval task, respectively.
It is acceptable to spend more time at the training phase to get the binary codes with better quality as the training phase is offline.
The query time of our method is comparable to the most baseline methods.
We notice that DJSRH consumes the least time at the query stage.
The main reason is that DJSRH adopts Alexnet as its basic network which is shallower than CNN-F.
Therefore, the time of feature extraction is shorter than that of other methods which use CNN-F as the feature extraction network.
\label{Experimental Result}
\begin{table}[t]
\caption{Performance comparison for three variants on MIR Flickr, NUS-WIDE and IAPR TC-12 on 32 bits.}
\vspace{-0.1cm}
\centering
\setlength{\tabcolsep}{2mm}{
\begin{tabular}{|c|c|c|c|c|c|c|}
\hline
\multirow{2}{*}{Methods} & \multicolumn{2}{c|}{MIR Flickr} & \multicolumn{2}{c|}{NUS-WIDE}& \multicolumn{2}{c|}{IAPR TC-12} \\
  \cline{2-7}
  \multirow{2}{*}{} &{I2T} & {T2I} & {I2T} & {T2I} & {I2T} & {T2I} \\\hline
TA-ADCMH-I	&	0.8699	&	0.8682	&	0.7658	&	0.7997	&	0.6213	&	0.6282	\\\hline
TA-ADCMH-II	&	0.8598	&	0.8631	&	0.6901	&	0.7107	&	0.6358	&	0.5929	\\\hline
TA-ADCMH-III&   0.8731	&	0.8827	&	0.7524	&	0.7912	&	0.6281	&	0.5995	\\\hline
TA-ADCMH	&	0.8758	&	0.9083	&	0.7892	&	0.8001	&	0.6391	&	0.6287	\\\hline
\end{tabular}} \label{lab:7}
\end{table}
\label{Experimental Results}
\subsection{Effects of task-adaptive hash function learning}
Our method is designed to learn task-adaptive hash functions by additionally regressing the query modality representation to the class label. With further semantic supervision, the query-specific modality representation can effectively capture the query semantics of different cross-modal retrieval tasks. To verify the effects of this part, we design two variant methods TA-ADCMH-I and TA-ADCMH-II for performance comparison.
1) TA-ADCMH-I directly performs semantic regression from class label to the shared hash codes instead of the query-specific modality representation.
 Mathematically, the optimization objective function of TA-ADCMH-I becomes
\begin{small}
\begin{equation*}
\label{eq:smm}
\begin{split}
\min \limits_{\mathbf{B},\theta_{x},\theta_{y},\textbf{V}}-&\sum\limits_{i,j=1}^n(S_{ij}\Phi_{ij}\,-\, \log(1+e^{\Phi_{ij}}))+\lambda\lVert{\textbf{B}-\textbf{F}\rVert}_F^2+\beta\lVert{\textbf{B}-\textbf{G}\rVert}_F^2\\
&+\mu\lVert{\textbf{B}-\textbf{V}\textbf{L}\rVert}_F^2+\nu\,(\lVert{\textbf{F}\textbf{1}\rVert}_F^2
+\lVert{\textbf{G}\textbf{1}\rVert}_F^2+\lVert{\textbf{V}\rVert}_F^2)\\
 &\quad s.t. \ \textbf{B}\in{\left \{-1,1 \right \}}^{r\times n}\\
\end{split}
\end{equation*}
\end{small}
The binary codes are calculated as $\textbf{B}=\texttt{sgn}(\lambda \textbf{F}+\beta \textbf{G}+\mu \textbf{V}\textbf{L})$, where $\lambda$, $\beta$, $\mu$ and $\nu$ are balance parameters. $\textbf{L}$ is the class label. $\textbf{V}$ is the projection matrix which regresses the hash codes $\textbf{B}$ to the semantic label $\textbf{L}$. The term $ \nu\,\lVert{\textbf{F}\textbf{1}\rVert}_F^2
+\nu\,\lVert{\textbf{G}\textbf{1}\rVert}_F^2\,(\textbf{1}=[1,...,1]^\texttt{T}\in \mathbb{R}^{n})$ is employed to keep the balance of each bit of hash codes for all the training points.
2) The other variant method TA-ADCMH-II performs the pair-wise semantic supervision without employing any semantic information. Mathematically, the optimization objective function of TA-ADCMH-II becomes
\begin{small}
\begin{equation*}
\label{eq:sm}
\begin{split}
\min \limits_{\mathbf{B},\theta_{z_1},\theta_{z_2}}-\sum\limits_{i,j=1}^n(S_{ij}\Phi_{ij}\,-\, \log(1+e^{\Phi_{ij}}))+&\lambda\lVert{\textbf{B}-\textbf{Z}_1\rVert}_F^2+\beta\lVert{\textbf{B}-\textbf{Z}_2\rVert}_F^2
+\nu\,(\lVert{\textbf{Z}_1\textbf{1}\rVert}_F^2
+\lVert{\textbf{Z}_2\textbf{1}\rVert}_F^2)\\
 &\quad s.t.  \  \textbf{B}\in{\left \{-1,1 \right \}}^{r\times n}\\
\end{split}
\end{equation*}
\end{small}
The binary codes are computed as $\textbf{B}=\texttt{sgn}(\lambda \textbf{Z}_1+\beta \textbf{Z}_2)$, where $\lambda$, $\beta$, $\nu$ and are balance parameters.
$\textbf{Z}_1$ and $\textbf{Z}_2$  represent the deep features of images and texts, respectively.
The performance comparison results of the two variant methods are shown in Table \ref{lab:7} on two sub-retrieval tasks.
The results demonstrate a fact that our method can outperform the variants TA-ADCMH-I and TA-ADCMH-II on three datasets for cross-modal retrieval.
These results prove that the task-adaptive hash function learning is effective on improving the cross-modal retrieval performance.
\subsection{Effects of discrete optimization}
We devise a variant method TA-ADCMH-III for comparison to validate the effect of discrete optimization.
Specifically, we utilize a relaxing strategy which first adopts continuous constraints instead of discrete constraints and then binarizes the real-valued solution into hash codes by thresholding. In Eq.(\ref{eq:W}) and Eq.(\ref{eq:w2}), we directly discard the constraints $\textbf{B}_1\in{\left \{-1,1 \right \}}^{r\times n}$ and $\textbf{B}_2\in{\left \{-1,1 \right \}}^{r\times n}$. Therefore, the relaxed hash codes can be calculated as
$\textbf{B}_1= (\lambda_1 \textbf{F}_1+\beta_1 \textbf{G}_1)/(\lambda_1+\beta_1)$ for I2T task, $\textbf{B}_2= (\lambda_2 \textbf{F}_2+\beta_2 \textbf{G}_2)/(\lambda_2+\beta_2)$ for T2I task. Table \ref{lab:7} shows the comparison results of TA-ADCMH-III and TA-ADCMH on MIR Flickr, NUS-WIDE, and IAPR TC-12, respectively. The results demonstrate that TA-ADCMH can achieve superior performance than TA-ADCMH-III, which further validates the quantization errors are minimized by the effect discrete optimization.

\label{sec:5.3}
\begin{figure*}[t]
\centering
\subfigure[Effects of $\lambda_1$ on mAP]{\includegraphics[scale=0.3]{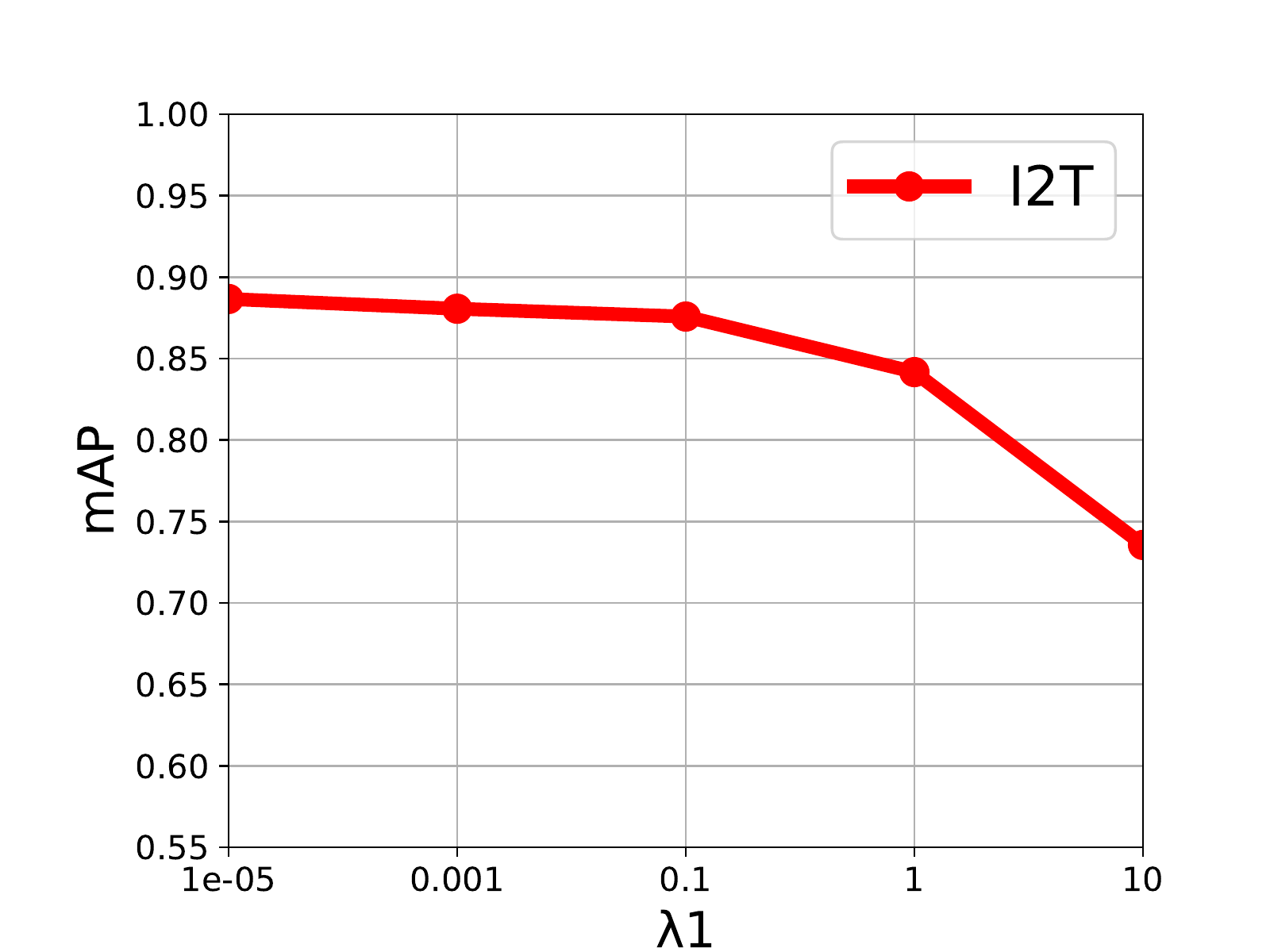}}
\subfigure[Effects of $\beta_1$ on mAP]{\includegraphics[scale=0.3]{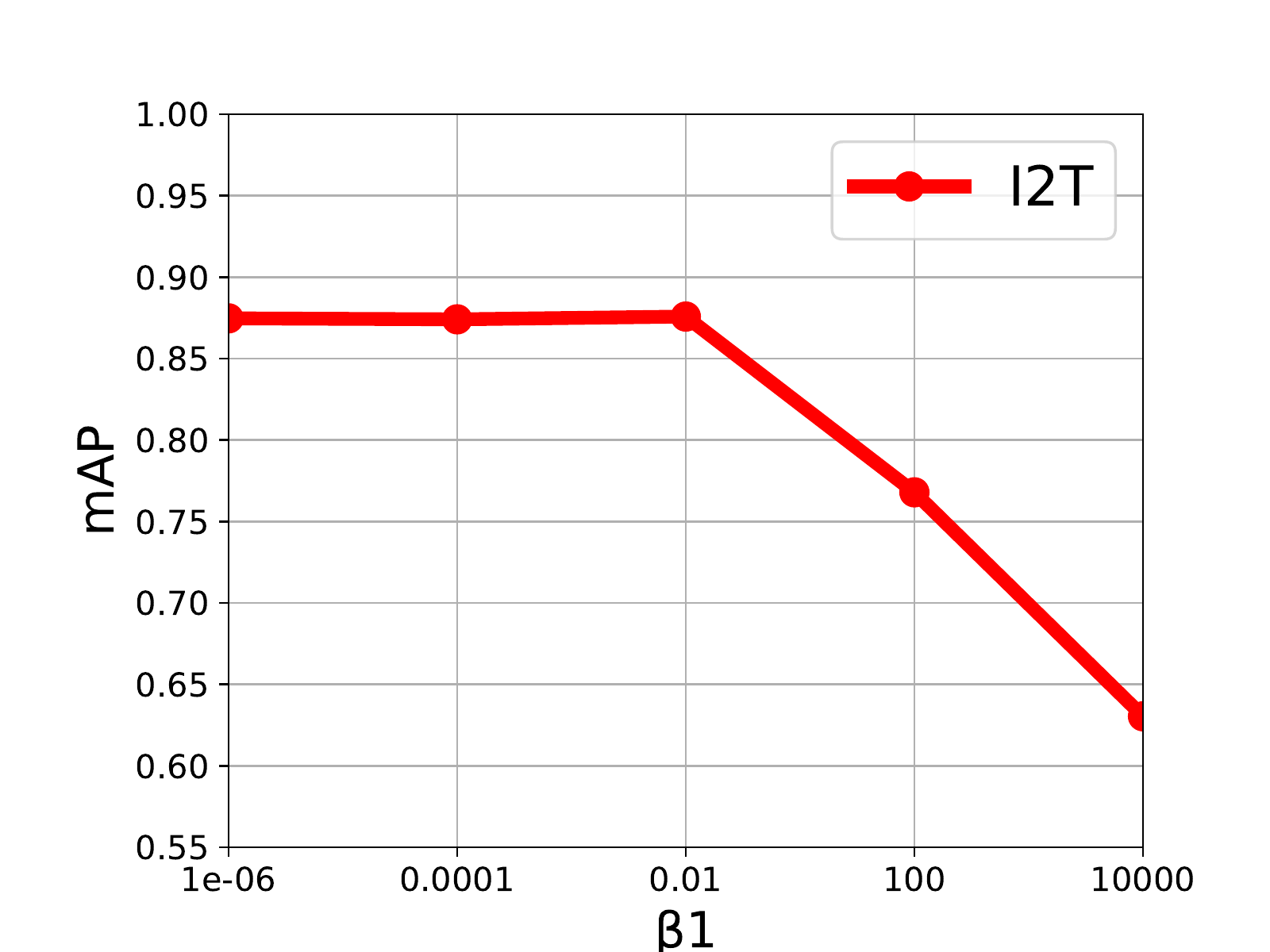}}
\subfigure[Effects of $\mu_1$ on mAP]{\includegraphics[scale=0.3]{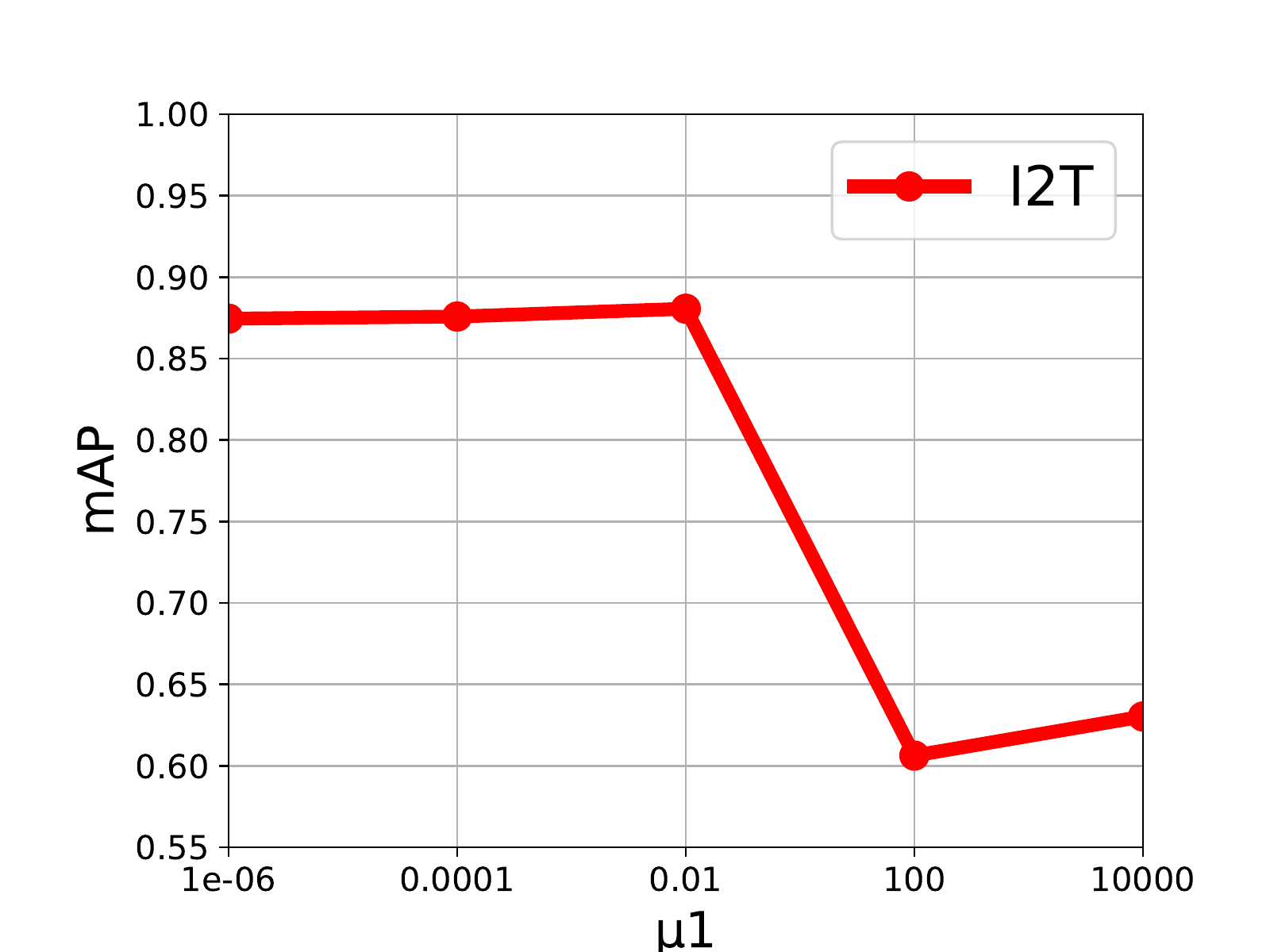}}
\subfigure[Effects of $\nu_1$ on mAP]{\includegraphics[scale=0.3]{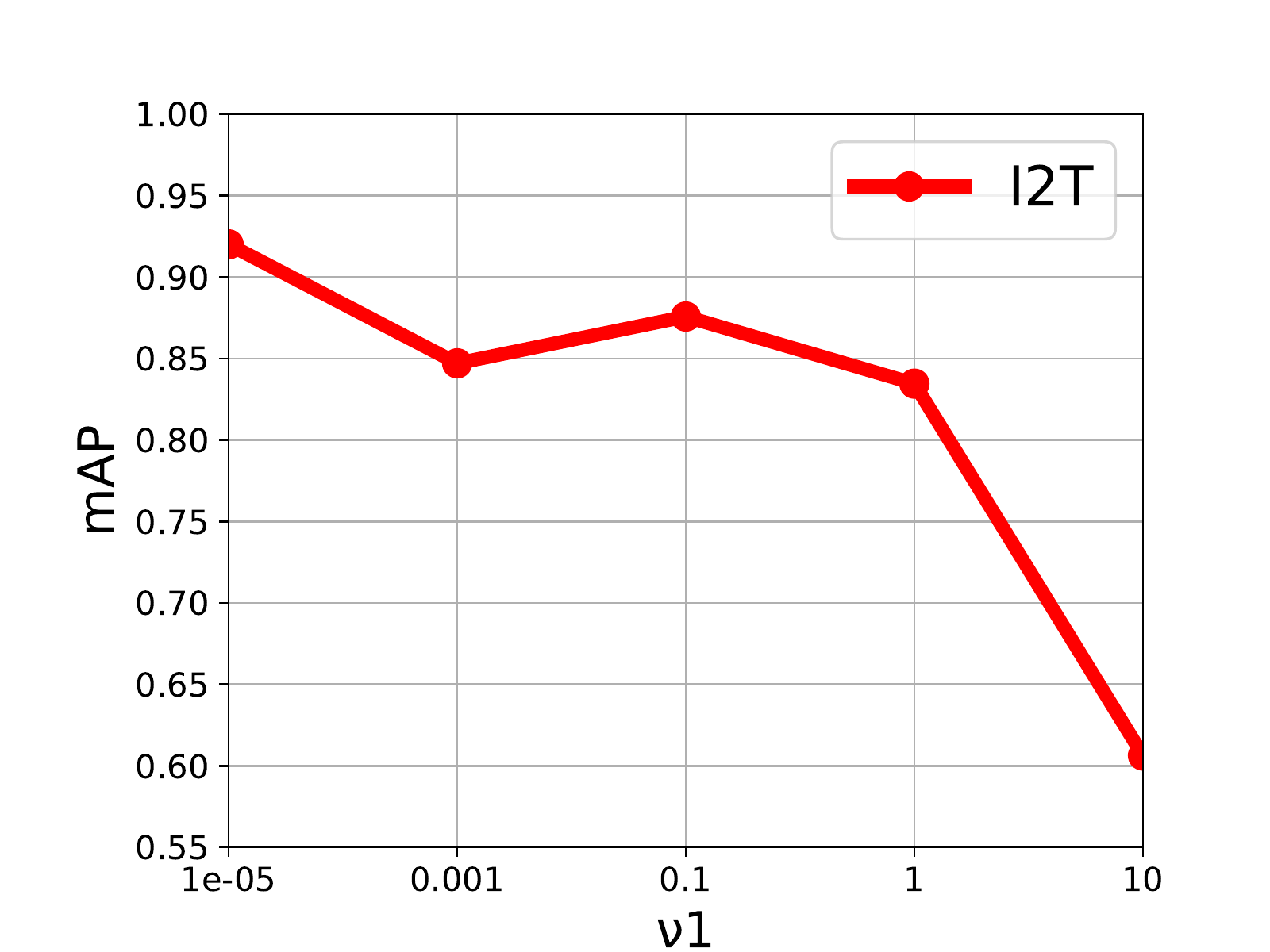}}
\vspace{-0.4cm}
\caption{Parameter experiments on I2T retrieval task. }
\label{fig:6}
\end{figure*}
\label{sec:5.4}
\begin{figure*}[t]
\centering
\subfigure[Effects of $\lambda_2$ on mAP]{\includegraphics[scale=0.3]{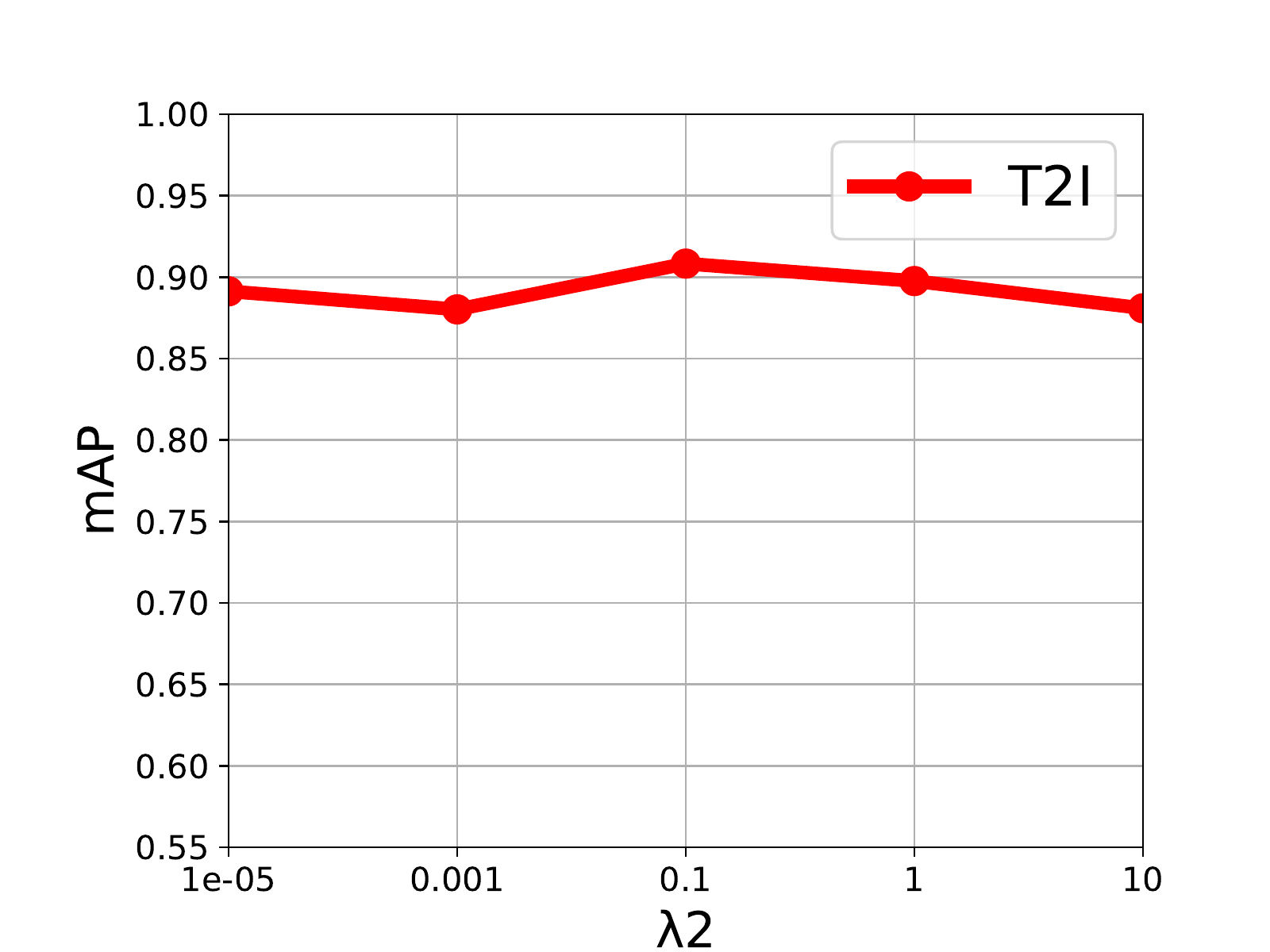}}
\subfigure[Effects of $\beta_2$ on mAP]{\includegraphics[scale=0.3]{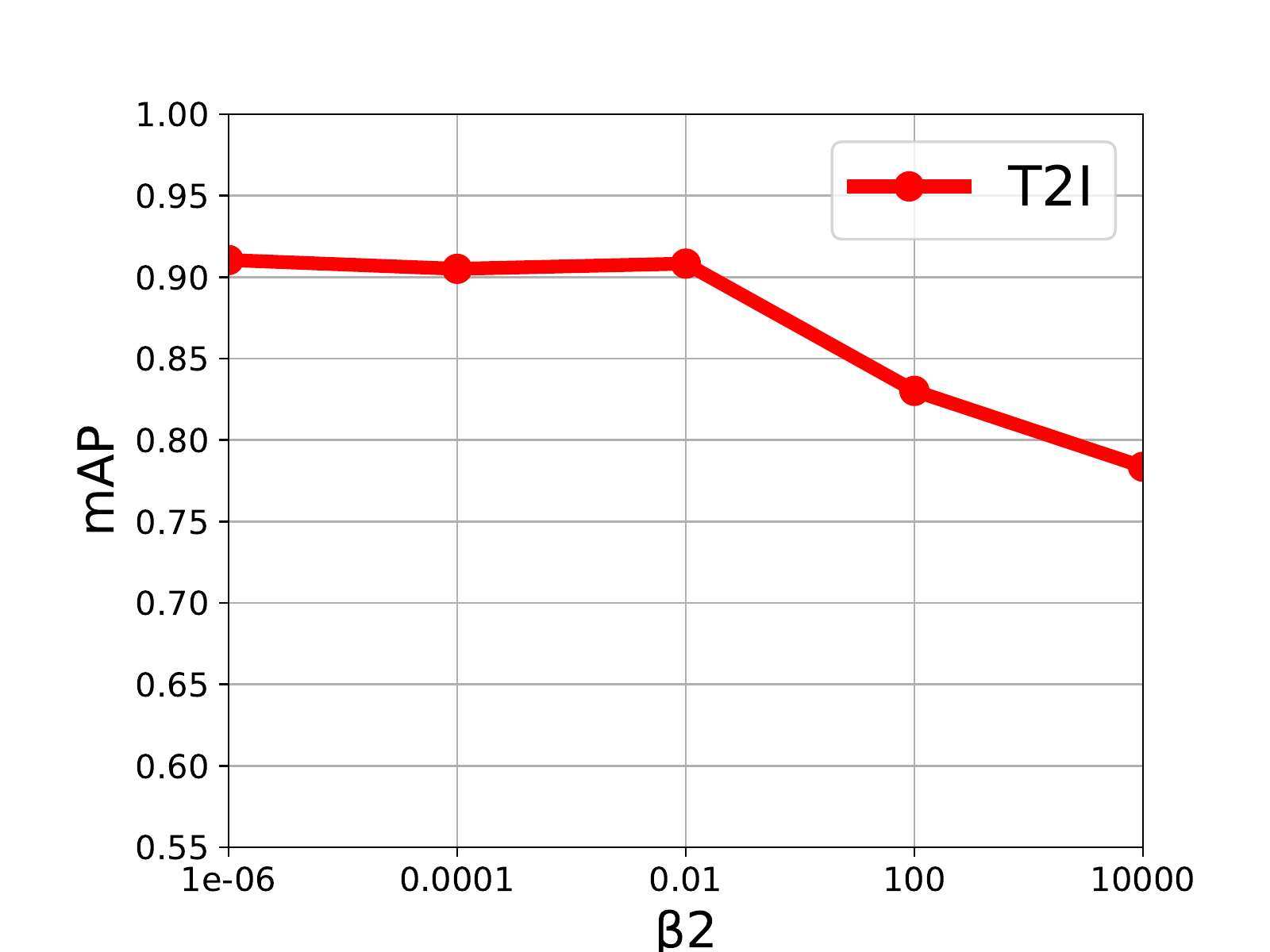}}
\subfigure[Effects of $\mu_2$ on mAP]{\includegraphics[scale=0.3]{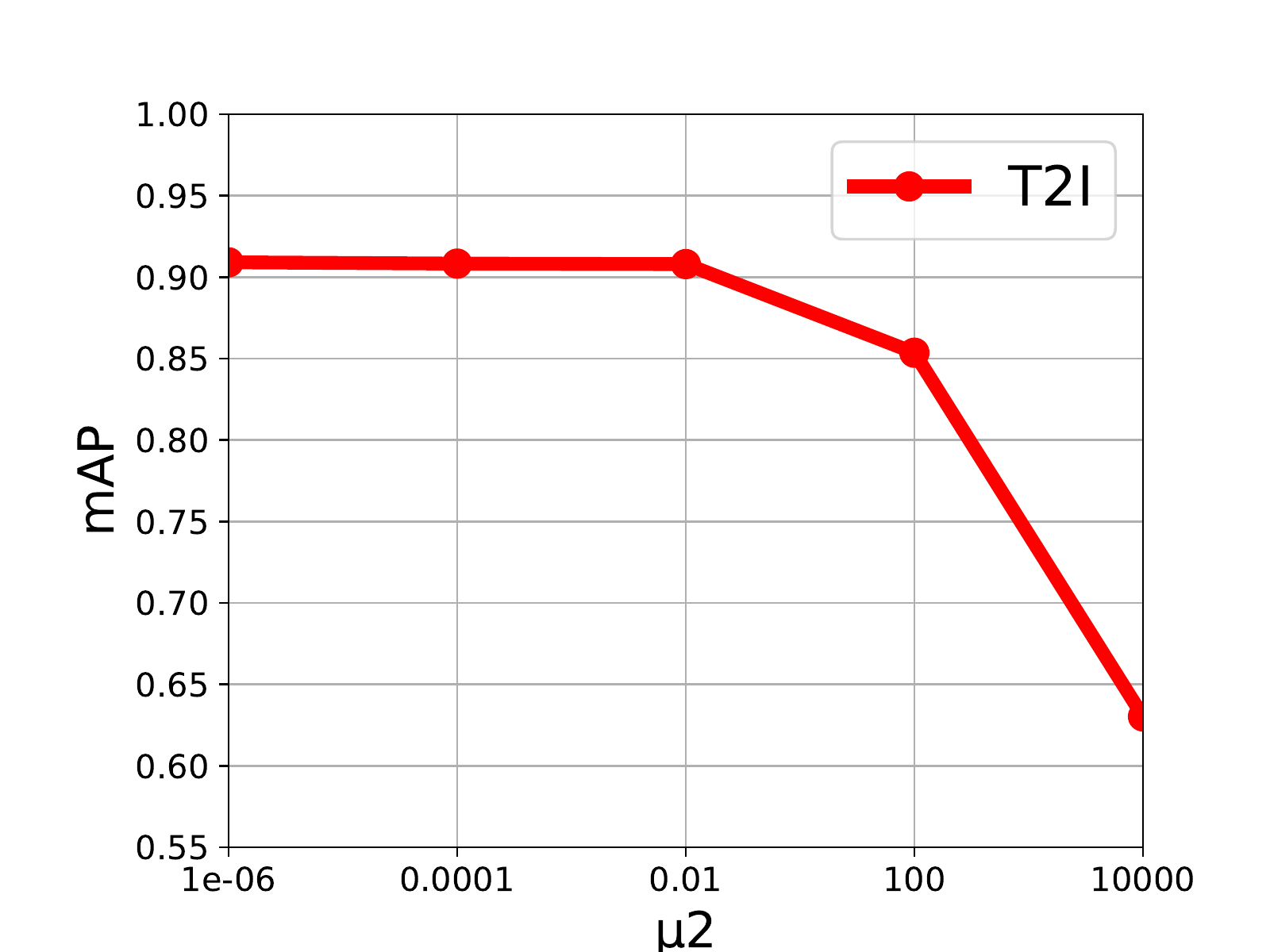}}
\subfigure[Effects of $\nu_2$ on mAP]{\includegraphics[scale=0.3]{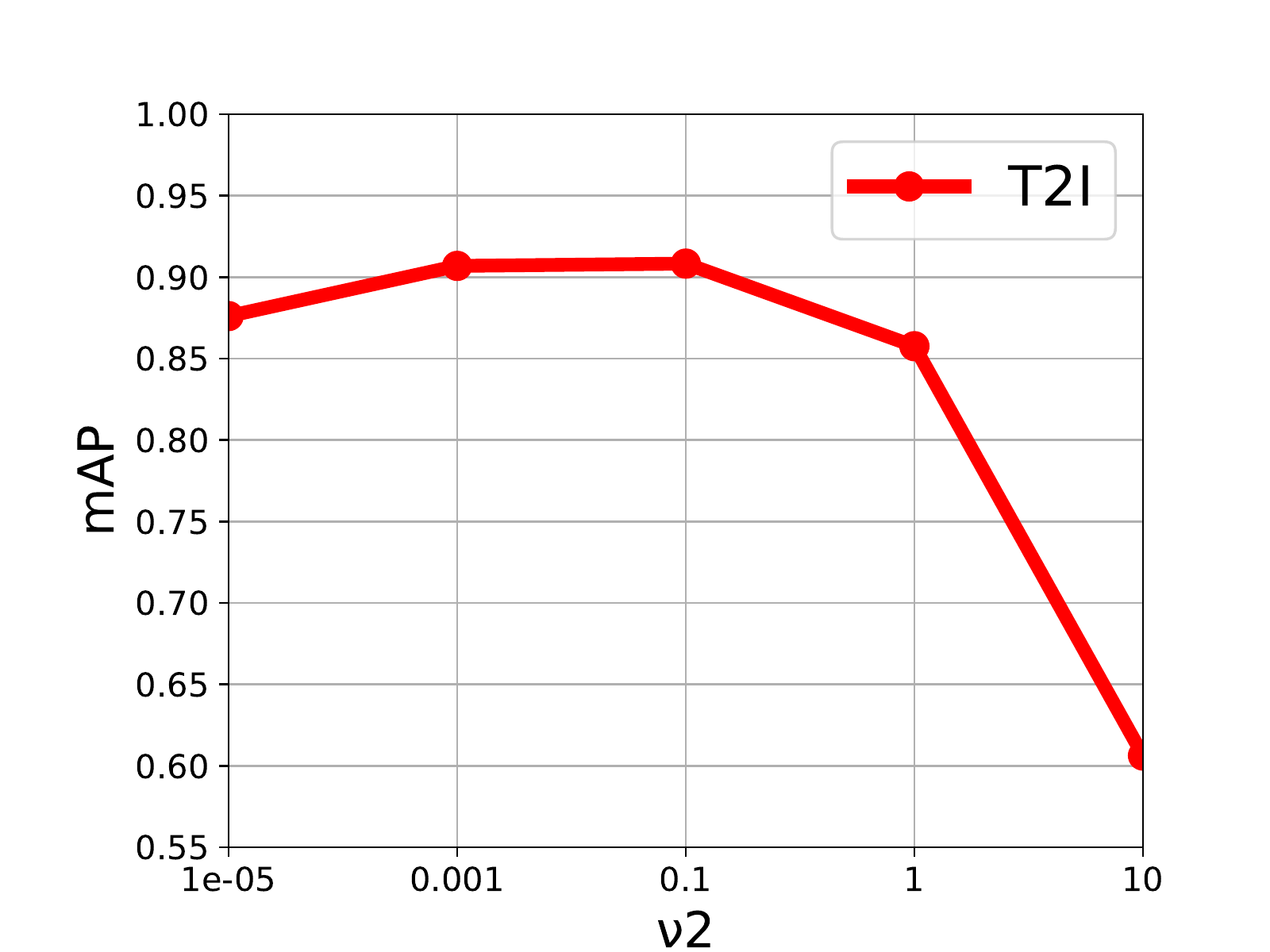}}
\vspace{-0.4cm}
\caption{Parameter experiments on T2I retrieval task . }
\label{fig:77}
\end{figure*}
\captionsetup[figure]{labelfont={bf},name={Figure},singlelinecheck=off}
\begin{figure*}[t]
\centering
\subfigure[]{\includegraphics[scale=0.23]{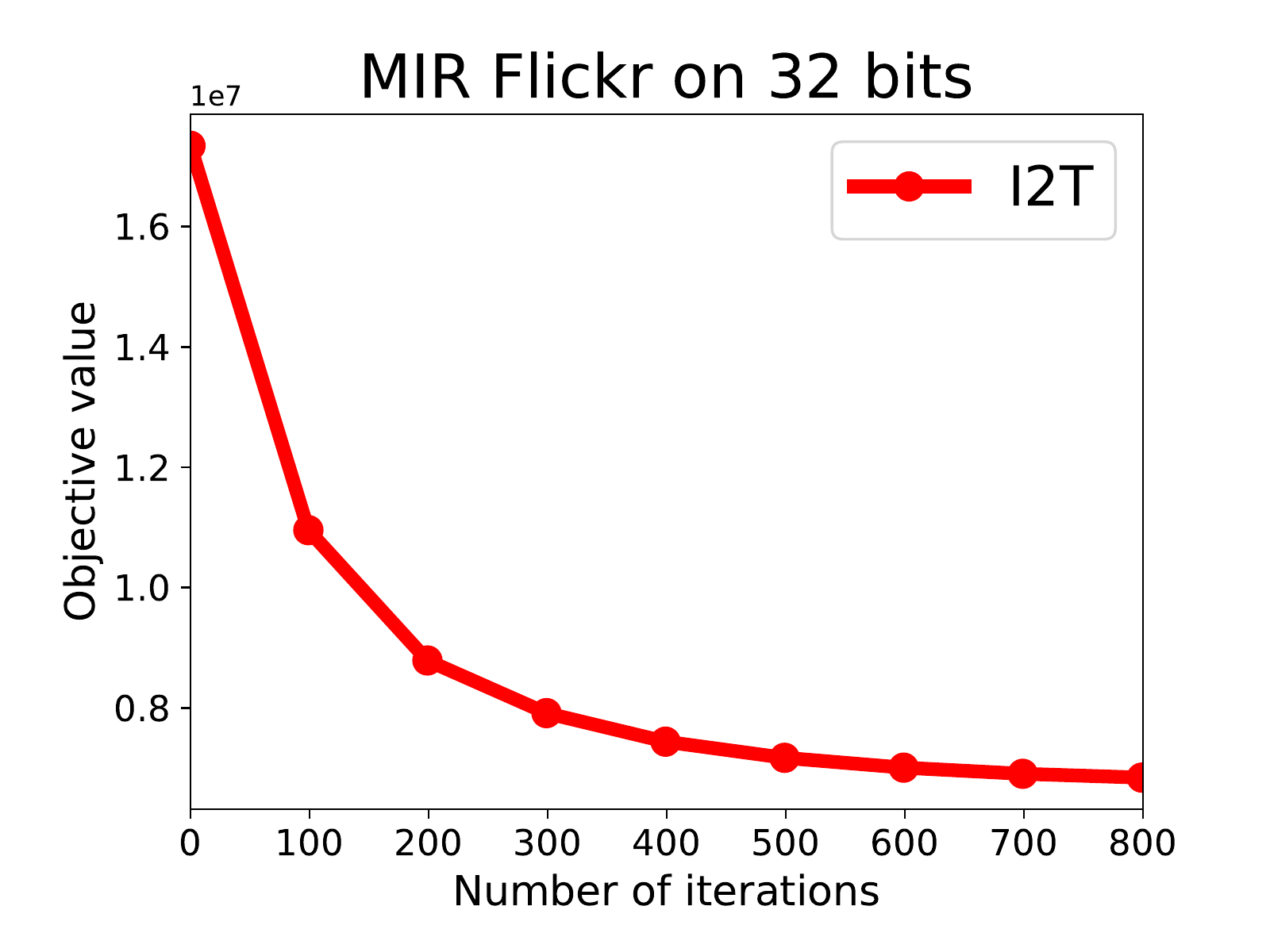}}
\subfigure[]{\includegraphics[scale=0.23]{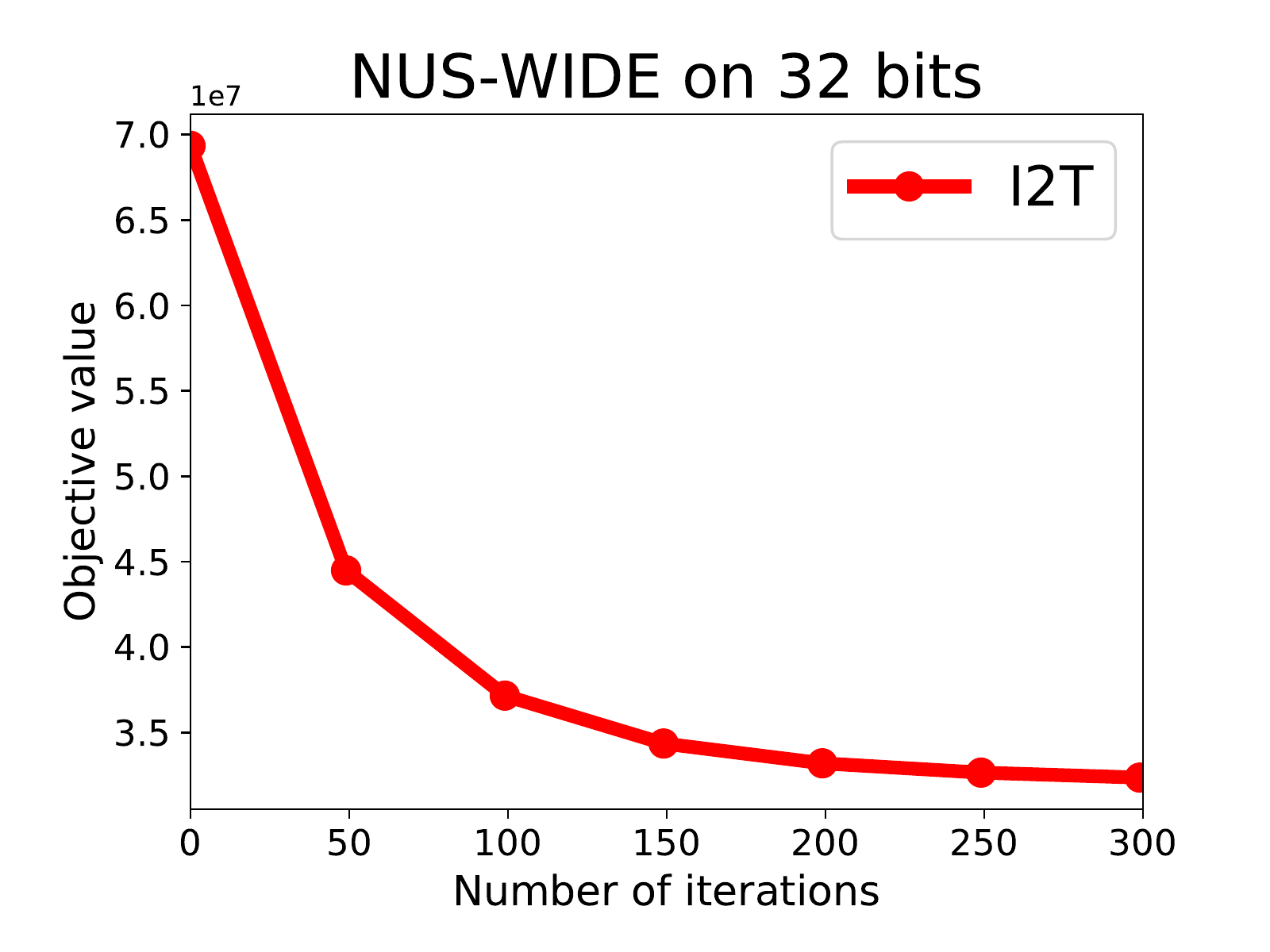}}
\subfigure[]{\includegraphics[scale=0.23]{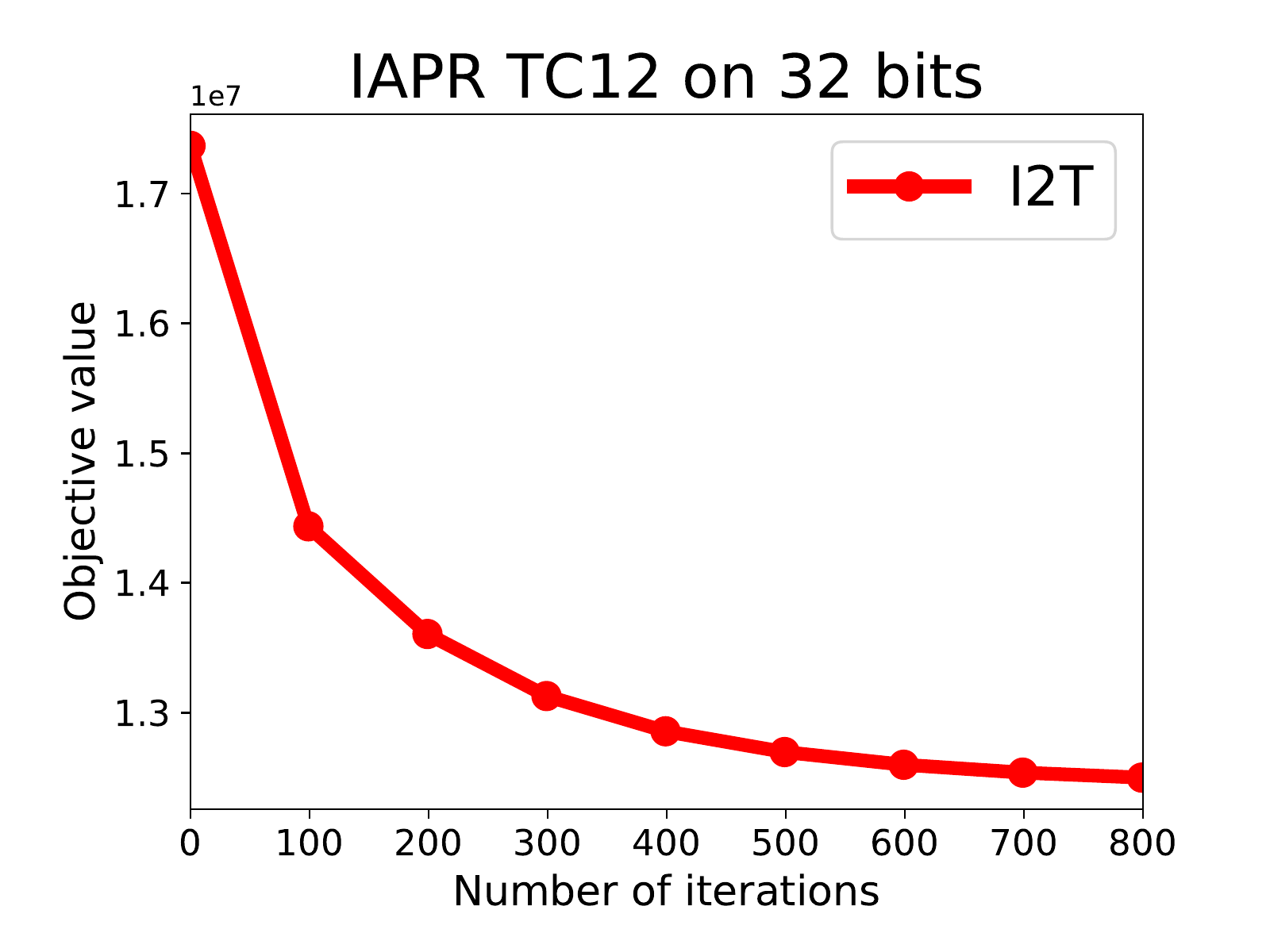}}\\
\subfigure[]{\includegraphics[scale=0.23]{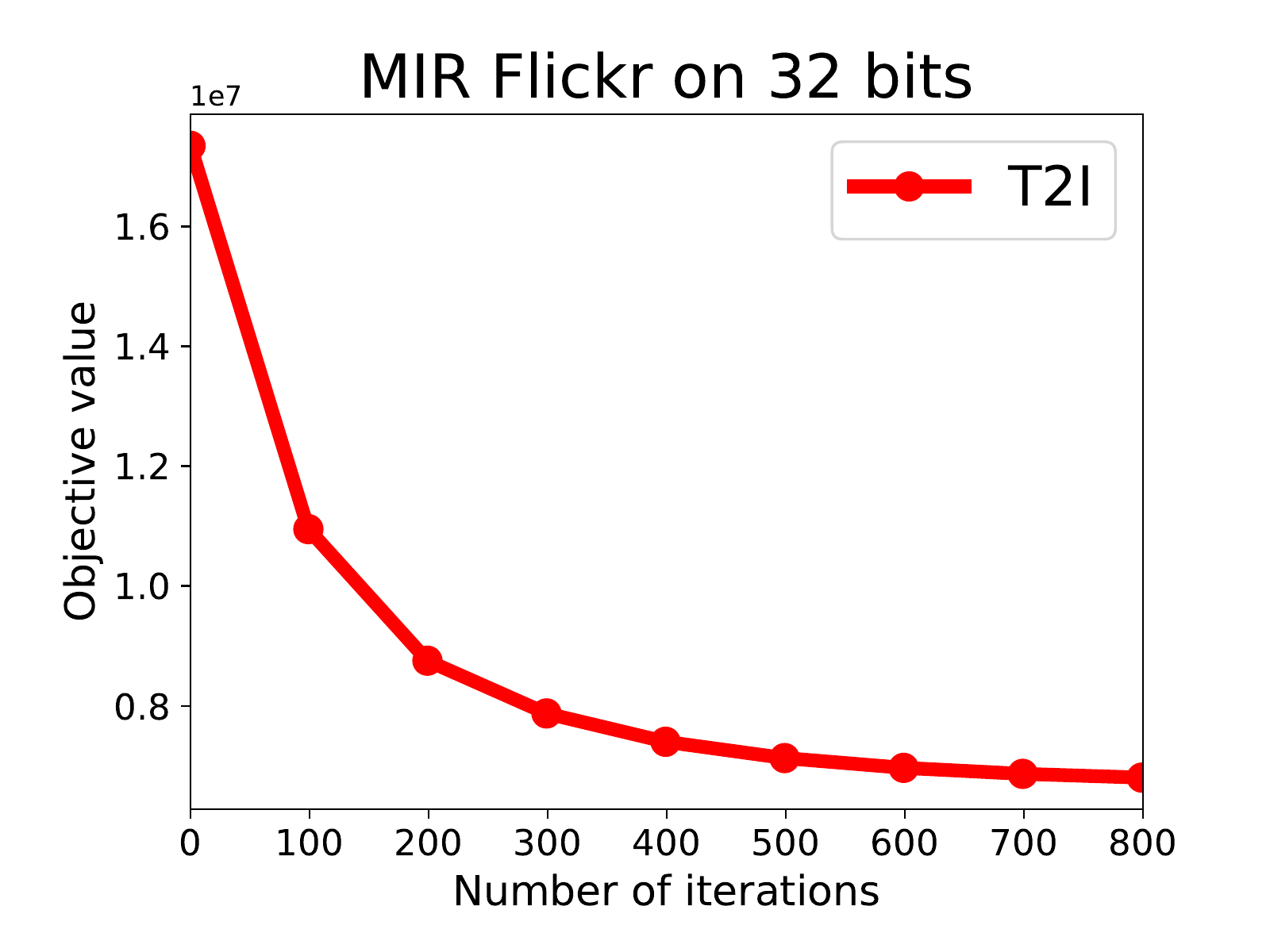}}
\subfigure[]{\includegraphics[scale=0.23]{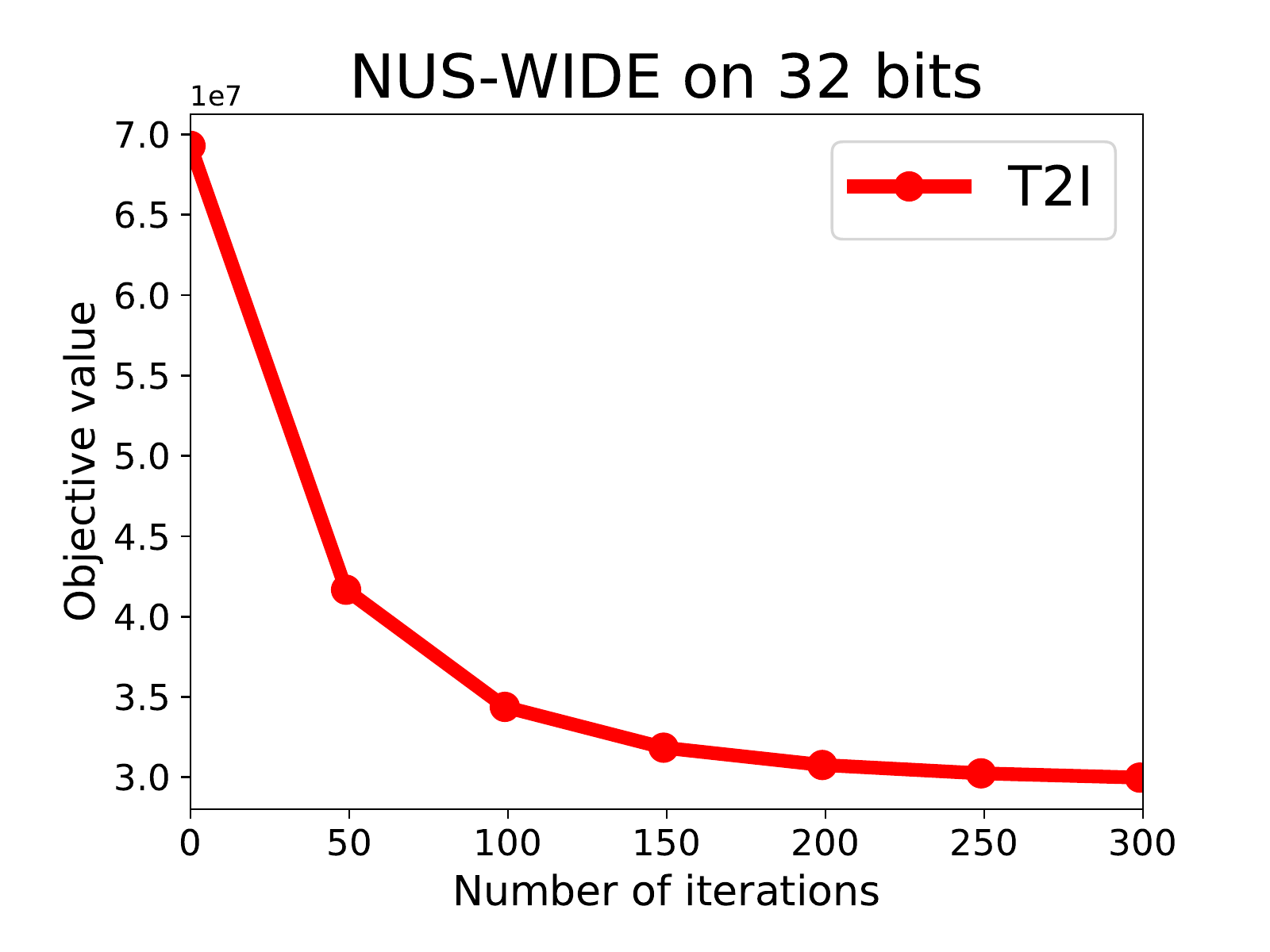}}
\subfigure[]{\includegraphics[scale=0.23]{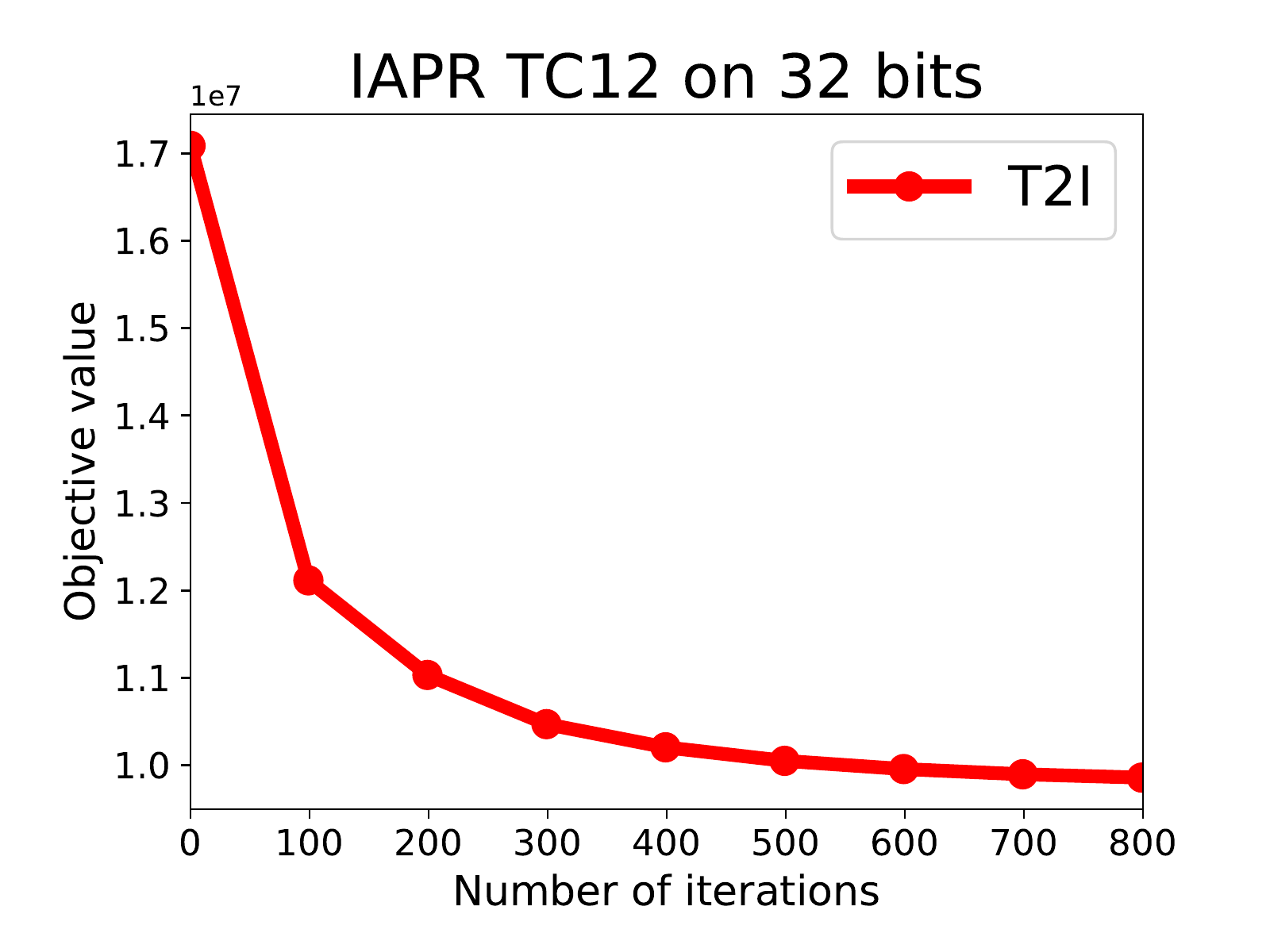}}
\vspace{-0.4cm}
\caption{Convergence curves on MIR Flickr, NUS-WIDE and IAPR TC-12.}
\label{fig:8}
\end{figure*}
\subsection{Parameter experiments}
The empirical analysis of parameter sensitivity are conducted on MIR Flickr with 32 bits.
Specifically, we report the mAP results with eight involved parameters for two different retrieval tasks.
In the I2T retrieval task, there are four involved parameters: $\lambda_1$, $\beta_1$, $\mu_1$ and $\nu_1$.
For $\lambda_1$ and $\nu_1$, we tune them from $ [10^{-5},10^{-3},10^{-1},1, 10]$ by fixing the other parameters.
In Figure \ref{fig:6} (a) and (d),
the performance is relatively stable when $\lambda_1$ and $\nu_1$ are in the range of $[10^{-5}, 10^{-1}]$ and $[10^{-3}, 1]$ for I2T retrieval task, respectively.
The mAP performance decreases sharply as the two parameters varied from $1$ to $10$.
For the parameter $\beta_1$ and $\mu_1$, we tune them from $10^{-6}$ to $10^{5}$.
As shown in Figure \ref{fig:6} (b) and (c), the I2T retrieval task is indeed influenced by $\beta_1$ and $\mu_1$. Specifically, with $\beta_1$ increasing ($\beta_1$ $<$ $10^{-2}$), the mAP performance is relatively stable.
However, when $\beta_1 > 10^{-2}$, the performance degrades quickly.
Hence, $\beta_1$ can be chosen within the range of $[10^{-6}, 10^{-2}]$.
Moreover, it is capable of observation from Figure \ref{fig:6} (c) that the performance is also relatively stable with $\mu_1$  in the range of $[10^{-6}, 10^{-2}]$.
The mAP values decreases quickly when $\mu_1$ is larger than $10^{-2}$.
In particular, $\mu_1$ can be chosen from the range between $[10^{-6}, 10^{-2}]$.
The remain parameters $\lambda_2$, $\beta_2$, $\mu_2$ and $\nu_2$ are involved for T2I retrieval task.
For $\lambda_2$ and $\nu_2$, we tune them from $[10^{-5},10^{-3},10^{-1},1, 10]$
by fixing the other parameters. For $\beta_2$ and $\mu_2$, we tune them from $10^{-6}$ to $10^{5}$.
The detailed explanation of results are shown in Figure \ref{fig:77}.
From Figure \ref{fig:77} (a), (c), (d), we have a conclusion that the mAP results of T2I task
are in a steady trend when $\lambda_2$ ranges from $10^{-5}$ to $ 10$, $\mu_2$ ranges from $10^{-6}$ to $ 10^{-2}$ and $\nu_2$ ranges from $10^{-5}$ to $10^{-1}$.
From Figure \ref{fig:77} (b), it can be discovered that the best performance of T2I task can be obtained when $\beta_2$ is in $[10^{-6}, 10^{-2}]$, and its performance decreases when $\beta_2$ is larger than $10^{-2}$.
In general, we can reach the conclusion that the parameters of TA-ADCMH are of vital importance to our experiments and the performance can be stable when the parameters are set within a reasonable range of values.
\subsection{Convergence analysis}
To analyze the convergence of TA-ADCMH, we display the convergence curves in Figure \ref{fig:8} with 32 bits on MIR Flickr, NUS-WIDE, and IAPR TC-12. For I2T retrieval task, the convergence analysis results of Eq.(\ref{eq:W}) are recorded in Figure \ref{fig:8} (a), (b) and (c). For T2I retrieval task, the convergence analysis results of the formula Eq.(\ref{eq:w2}) are recorded in Figure \ref{fig:8} (d), (e) and (f). In all the figures, the abscissa is the iteration numbers and the ordinate is the value
of the objective function. As shown in the figures,
we can find that TA-ADCMH achieves a stable minimum within 500 iterations for I2T task and T2I task on MIR Flickr dataset. We can also find that it converges within 300 iterations for both two retrieval tasks on NUS-WIDE dataset. On IAPR TC-12 dataset, it converges within 500 iterations for both two retrieval tasks.
The experimental results confirm that TA-ADCMH can converge gradually.
\subsection{Performance variations with training size}
\begin{figure*}[t]
\centering
\subfigure[]{\includegraphics[scale=0.35]{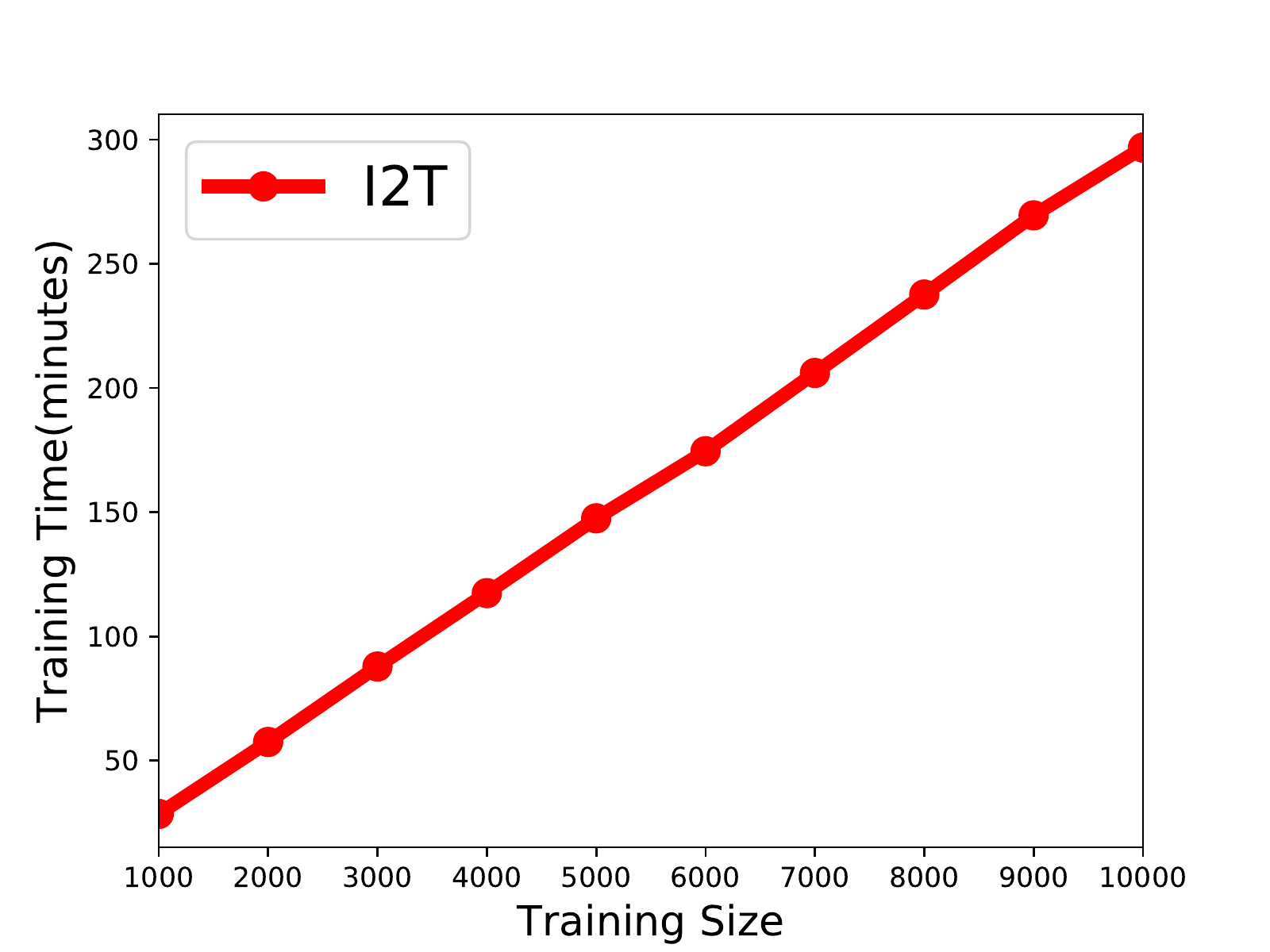}}
\subfigure[]{\includegraphics[scale=0.35]{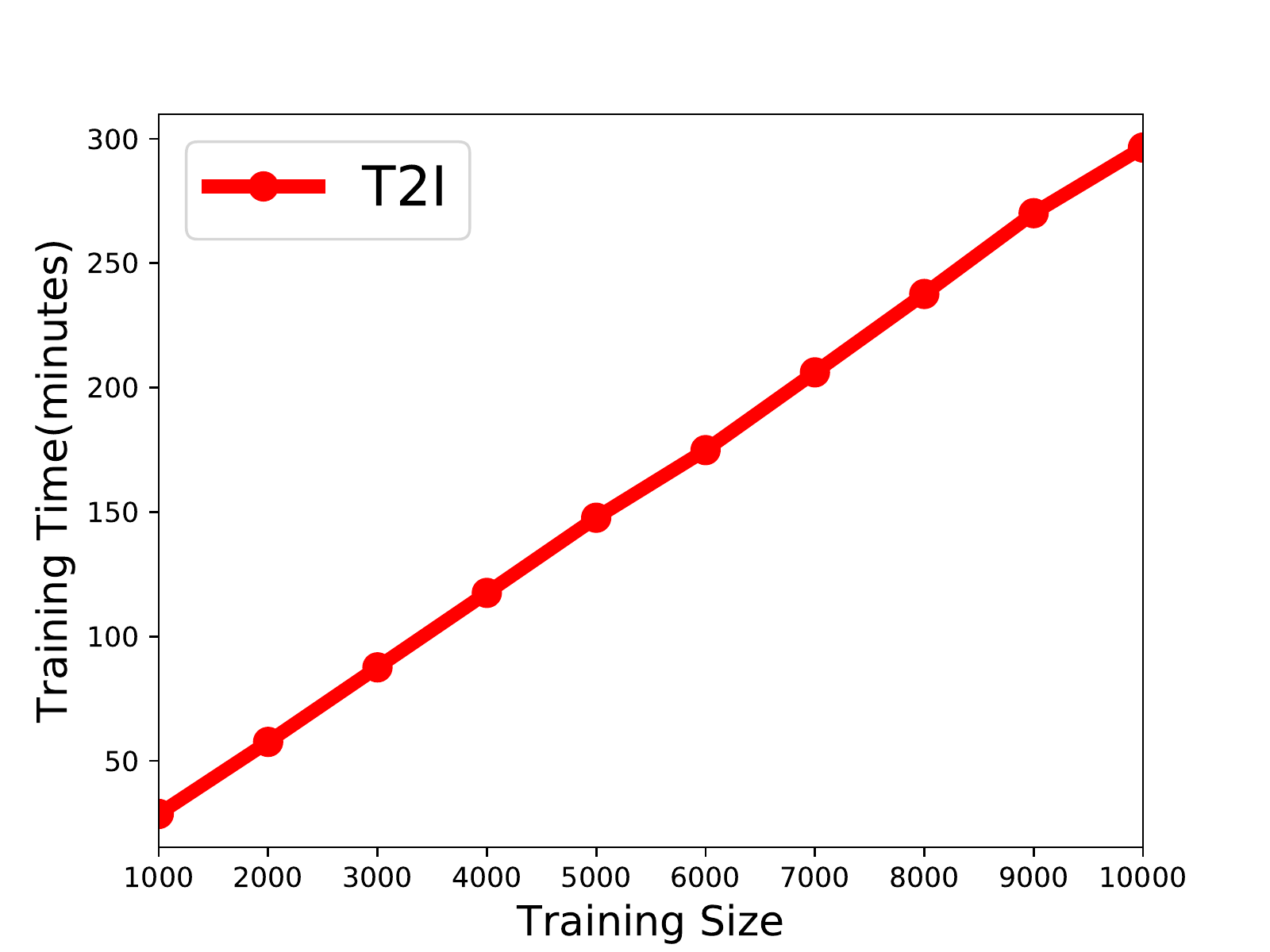}}
\vspace{-0.4cm}
\caption{Variations of training time with training data size on NUS-WIDE.}
\label{trainsizeFig}
\end{figure*}

We evaluate the performance variations of our proposed TA-ADCMH with respect to the training size on NUS-WIDE in this subsection.
The results are shown in Table \ref{trainsizeTab} and Figure \ref{trainsizeFig} when the hash code length is fixed to 32 bits.
From Figure \ref{trainsizeFig}, we can observe that the training time of TA-ADCMH increases linearly with the training size on two sub-retrieval task,
which validates the linear computational complexity of the proposed method.
From Table \ref{trainsizeTab}, we can find that the performance of TA-ADCMH first increases with training data size
and then becomes stable after certain point (training size 6000).
The experimental phenomenon reveals that the proposed method can achieve satisfactory performance even with small training data.
\begin{table}[t]
\small
\centering
\caption{Performance variations with the training size on NUS-WIDE.}
\setlength{\tabcolsep}{1mm}{
\begin{tabular}{|c|c|c|c|c|c|c|c|c|c|c|c|}
\hline
\multicolumn{2}{|c|}{\#Trainsize}	&1K&2K&3K&4K&5K&6K&7K&8K&9K&10K	\\\hline
\multirow{2}{*}{mAP} & I2T&0.5308&0.5459 &0.6339 &0.6940&0.7242 &0.7724 &0.7823 &0.7737&0.7819&0.7892\\ \cline{2-12}
                     & T2I&0.6222&0.6920&0.7010 & 0.7203&0.7391&0.7500&0.7967&0.7917&0.8013&0.8001 \\ \hline
\end{tabular}}\label{trainsizeTab}
\end{table}
\section{Conclusion}
In this work, we propose a Task-adaptive Asymmetric Deep Cross-modal Hashing (TA-ADCMH) method. It learns task-adaptive hash functions for different cross-modal retrieval tasks. The deep learning framework jointly optimizes the semantic preserving from multi-modal deep representations to the hash codes, and the semantic regression from the query-specific representation to the explicit labels. The learned hash codes can effectively preserve the multi-modal semantic correlations, and meanwhile, adaptively capture the query semantics. Further, we devise a discrete optimization scheme to effectively solve the discrete binary constraints of binary codes.
Experimental results on three public cross-modal retrieval datasets validate the superiority of the proposed method.

\section{Acknowledgements}
This work was supported in part by the National Natural Science Foundation of China under Grant 61802236, Grant 61772322, and Grant U1836216, in part by the Natural Science Foundation of Shandong, China, under Grant ZR2020YQ47 and Grant ZR2019QF002, in part by the Major Fundamental Research Project of Shandong, China, under Grant ZR2019ZD03, in part by the Youth Innovation Project of Shandong Universities, China, under Grant 2019KJN040, and in part by the Taishan Scholar Project of Shandong, China, under Grant ts20190924.


\bibliography{mybibfile}

\end{document}